\documentclass[aps,amsfonts,preprint]{revtex4}
\usepackage{epsfig,amsmath,amssymb,bm,epsf,graphics,psfrag,color,ulem}

\def\Longarrow{\protect\@lra}
\def\@lra{\relbar\joinrel\relbar\joinrel\relbar\joinrel%
          \relbar\joinrel\rightarrow}

\def\inf{\infty}

\def\myOmega{\Omega}

\begin{document}
\title{Emergence of  $\bm{h/e}$-period oscillations in the critical temperature of
small superconducting rings threaded by magnetic flux}
\author{Tzu-Chieh Wei}
\affiliation{Institute for Quantum Computing and Department of Physics and
Astronomy, University of Waterloo, Waterloo, ON N2L 3G1, Canada}
\author{Paul M. Goldbart}
\affiliation{Department of Physics and Frederick Seitz Materials Research Laboratory,\\
University of Illinois at Urbana-Champaign, Urbana, Illinois 61801, U.S.A.}

\begin{abstract}
 As a function of the magnetic flux threading the
object, the Little-Parks oscillation in the critical temperature of a
large-radius, thin-walled superconducting ring or hollow cylinder has a period
given by $h/2e$, due to the binding of electrons into Cooper pairs.  On the
other hand, the single-electron Aharonov-Bohm oscillation in the resistance or
persistent current for a clean (i.e.~ballistic) normal-state system having the
same topological structure has a period given by $h/e$.  A  basic question is
whether the Little-Parks oscillation changes its character, as the radius of
the superconducting structure becomes smaller, and even comparable to the
zero-temperature coherence length. We supplement a physical argument that the
$h/e$ oscillations should also be exhibited with a microscopic analysis of
this regime, formulated in terms of the Gor'kov approach to BCS theory.  We
see that, as the radius of the ring is made smaller, an oscillation in the
critical temperature of period $h/e$ emerges, in addition to the usual
Little-Parks $h/2e$-period oscillation.  We argue that in the clean limit
there is a superconductor-normal transition at nonzero flux, as the ring
radius becomes sufficiently small, and that the transition can be either
continuous or discontinuous, depending on the radius and the external flux. In
the dirty limit, we argue that the transition is rendered continuous, which
results in continuous quantum phase transitions tuned by flux and radius.
\end{abstract}

\pacs{74.62.-c,74.78.Na} \maketitle
\section{Introduction}
The Little-Parks critical-temperature oscillations, with magnetic flux, of a
large-radius thin-walled hollow cylindrical superconductor display a period
$h/2e$~\cite{LittleParks,Tinkham}. This oscillation period reflects the
binding of electrons into Cooper pairs~\cite{Schrieffer}. Recent experiments
by Liu et al.~\cite{PennState} have probed the regime in which the diameter of
the hollow cylinder is comparable to zero-temperature coherence of the
superconductor. Their results confirmed the prediction by de
Gennes~\cite{deGennes} of the destruction of superconductivity in small rings
for a certain regime of the external flux. Liu et al. \ raised the interesting
and fundamental issues of what would happen if the circumference of the
structure were to be smaller than the superconducting coherence length, as
well as whether or not the Ginzburg-Landau approach would be valid in this
regime. These issues make a microscopic treatment desirable.

On the other hand, the single-electron Aharonov-Bohm oscillations in the
resistance or persistent current in a clean metallic ring have period
$h/e$~\cite{ABmetal}. This leads to a related fundamental issue: As the radius
of the ring becomes smaller, how would the single-particle $h/e$ period
manifest itself in a Little-Parks type of experiment? Furthermore, does
disorder affect the oscillation of critical temperature and the character of
the transition between superconducting and normal states? And if so, how?
These questions are not only of theoretical interest, but are also likely to
be addressed experimentally, in view of recent progress in fabrication and
experiments on small superconducting
rings~\cite{BluhmKoshnickHuberMoler06,BluhmKoshnickHuberMoler07}.

Recent work by Czajka et al.~\cite{CzajkaMaskaMierzejewskiSledz05}, involing
the exact diagonalization of the Hubbard model for small numbers of sites and
the numerical solution of the Bogoliubov-de Gennes (BdG) equation, shows that
impurities can play an important role if they are located such that pinned
density waves are in phase with one another: charge-density-wave (CDW) order
would then be enhanced and superconducting order reduced. They also found that
the mean-field results obtained via the BdG equations are consistent with the
exact diagonalization results, even in the small systems they studied. Very
recently, a numerical study by Loder et al.~\cite{Dwave} and analytical work
by Juricic et al.~\cite{JuricicHerbutTesanovic07} and by
Barash~\cite{Barash07} on clean $d$-wave superconducting loops has shown
$h/e$-period oscillations in the supercurrent.

Ginzburg-Landau (GL) theory is valid near the superconducting transition, as
it is an expansion in powers of the superconducting order parameter. Therefore
it may not be applicable at sufficiently low temperatures. Moreover, the GL
approach cannot give a complete account of multi-connected geometries of small
size, as it is a description of the center-of-mass wavefunction of the Cooper
pairs. For small rings, in addition to propagating around the circumference
together, electrons in Cooper pairs can split apart and rejoin, and this
process is not included in the GL description. In this Paper, we study the
oscillations of the critical temperature of $s$-wave superconducting rings
theoretically, via consideration of the microscopic BCS theory of
superconductivity~\cite{BCS}, analyzed using Gor'kov's approach~\cite{Gorkov}.
Our central purpose is to address the issues raised by Liu et
al.~\cite{PennState} and mentioned above. Our focus is on the correction to
the oscillations that is due to finiteness of the radius of the
superconducting structure. We consider both the clean and dirty regimes, and
in the latter regime we shall ignore any tendency towards CDW ordering by
assuming that the significant configurations of the impurities are
sufficiently random
 that any pinned density waves are not in phase with one another. We shall see the
emergence of an $h/e$ oscillation in the critical temperature, as the radius
becomes smaller, and we shall also see that the transition to the normal state
can be either discontinuous or continuous in the clean limit but is always
continuous in the dirty limit.

The origin of the $h/2e$ oscillations lies in the fact that a Cooper pair
carries charge $2e$, which when circumnavigating the ring acquires a phase
$e^{i 2e \Phi/c\hbar}$, where $\Phi$ is the magnetic flux, linking the ring.
The period in flux is thus $h/2e$ (with $c$ conveniently set to
unity~\cite{footnotehe}). The emergence of $h/e$ oscillations for small rings
is due to the additional process in which electrons in a Cooper pair can, from
time to time (so to speak), separate, propagate separately, and rejoin, the
two trajectories having a nonzero winding number relative to one another. This
process, which can only occur for ring sizes comparable to or smaller than the
Cooper-pair size,  induces an oscillation with period $h/e$.

The organization of this Paper is as follows. In Sec.~\ref{sec:clean} we
discuss the critical-temperature oscillations in the clean limit and in
Sec.~\ref{sec:disordered} we include the effect of disorder and discuss the
dirty limit. We make some concluding remarks in Sec.~\ref{sec:conclude}. In
Appendix~\ref{app:Gorkov} we derive the effective one-dimensional Gor'kov
equations by averaging over the cross-section (or thickness) of the ring. In
Appendix~\ref{app:Detail} we provide supplementary details of the calculations
that lead to the results in the main text. In Appendix~\ref{app:Heuristic} we
provide two heuristic arguments for the emergence of the $h/e$ period: (1) by
examining Cooper's problem on a ring, and (2) by using an instanton approach.
These two arguments lead to the physical picture described above, and
complement the Gor'kov Green-function approach adopted in the main text.
\section{Clean limit}
\label{sec:clean}
\begin{figure} \centerline{ \rotatebox{0}{
        \epsfxsize=7.0cm
        \epsfbox{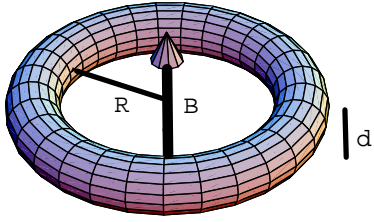}
} }
\vspace{-0.5cm}
 \caption{A ring with a magnetic flux threading through it. The radius of the ring is $R$ and the thickness in the cross-section is $d$. } \label{fig:ring}
\end{figure}
\subsection{Gor'kov equations}
We consider a ring (to be more precise, a torus) of radius $R$ (and thickness
$d$, namely, the diameter in the cross-section, which is smaller than both $R$
and the Cooper-pair size $\xi_0$) with a magnetic flux density $B$ threading
the ring parallel to the ring axis; see Fig.~\ref{fig:ring}. We can describe this field by the vector
potential $\vec{A}(r)= (B/2) \hat{z}\times \vec{r}$. On the ring itself (the
circumference of which is $L=2\pi R$), the vector potential $\vec{A}$ is given
by $\hat{\theta}\,{\Phi}/{L}$, where $\hat{\theta}$ is the azimuthal unit
vector and $\Phi$ is the total flux enclosed by the ring, i.e., $\Phi=\int
da\cdot B(\vec{r})$.

The normal and anomalous Green functions obey the Gor'kov equations~\cite{Gorkov,FetterWalecka,Schrieffer}
\begin{subequations}
\label{eqn:Gorkov}
\begin{eqnarray}
&&[i\hbar\omega_n-\frac{1}{2M}(-i\hbar\nabla-\frac{e
\vec{A}}{c})^2+\mu]G(\vec{r},\vec{r}';\omega_n)+\Delta(\vec{r})F^\dagger(\vec{r},\vec{r}';\omega_n)=
\hbar\delta(\vec{r}-\vec{r}'), \\
&&[-i\hbar\omega_n-\frac{1}{2M}(i\hbar\nabla-\frac{e
\vec{A}}{c})^2+\mu]F^\dagger(\vec{r},\vec{r}';\omega_n)-\Delta^*(\vec{r})G(\vec{r},\vec{r}';\omega_n)=
0,
\end{eqnarray}
\end{subequations}
where $M$ is the electron mass, $e$ is the electron charge, $\vec{r}$ and
$\vec{r}'$ are three-dimensional coordinates in the ring, and the order
parameter $\Delta$ is defined self-consistently via
\begin{equation}
\label{eqn:selfconsistency0} \Delta^*(\vec{r})=\frac{V}{\beta}\sum_{\omega_n}
F^\dagger(\vec{r},\vec{r}';\omega_n),
\end{equation}
in which $\omega_n\equiv 2\pi T(n+1/2)$ are  Matsubara frequencies,
$\beta\equiv 1/k_B T$, $T$ is the temperature, and $V$ is the BCS pairing
strength.

We now invoke the narrowness of the ring to justify dropping all dependence on
$\vec{r}$ and $\vec{r}'$ except that associated with the one-dimensional
coordinates along the ring, $x$ and $x'$. Owing to the physical periodicity of
the ring, all functions of $x$ and $x'$ are periodic with period $L$ (and the
vector potential is a constant along the circumference of the ring). The
Gor'kov equations then become one dimensional~\cite{Average}:
\begin{subequations}
\label{eqn:Gorkov1D}
\begin{eqnarray}
&&\Big[+i\hbar\omega_n-\frac{1}{2M}\big(i\hbar\partial_x+\frac{e \Phi}{c
L}\big)^2+\mu\Big]G(x,x';\omega_n)+\Delta({x})F^\dagger(x,x';\omega_n)=
\hbar\delta(x-x'), \\
&&\Big[-i\hbar\omega_n-\frac{1}{2M}\big(i\hbar\partial_x-\frac{e \Phi}{c
L}\big)^2+\mu\Big]F^\dagger(x,x';\omega_n)-\Delta^*(x)G(x,x';\omega_n)=
0.
\end{eqnarray}
\end{subequations}

\begin{figure}\vspace{0.5cm} \centerline{ \rotatebox{0}{
        \epsfxsize=7.0cm
        \epsfbox{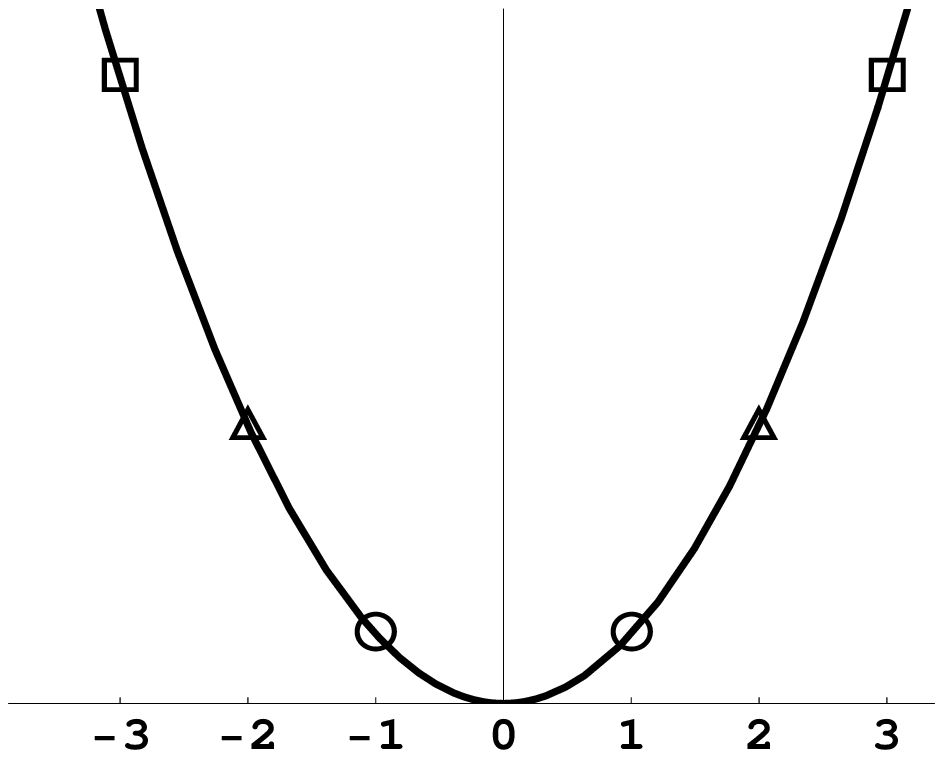}
} }
 \centerline{ \rotatebox{0}{
        \epsfxsize=7.0cm
        \epsfbox{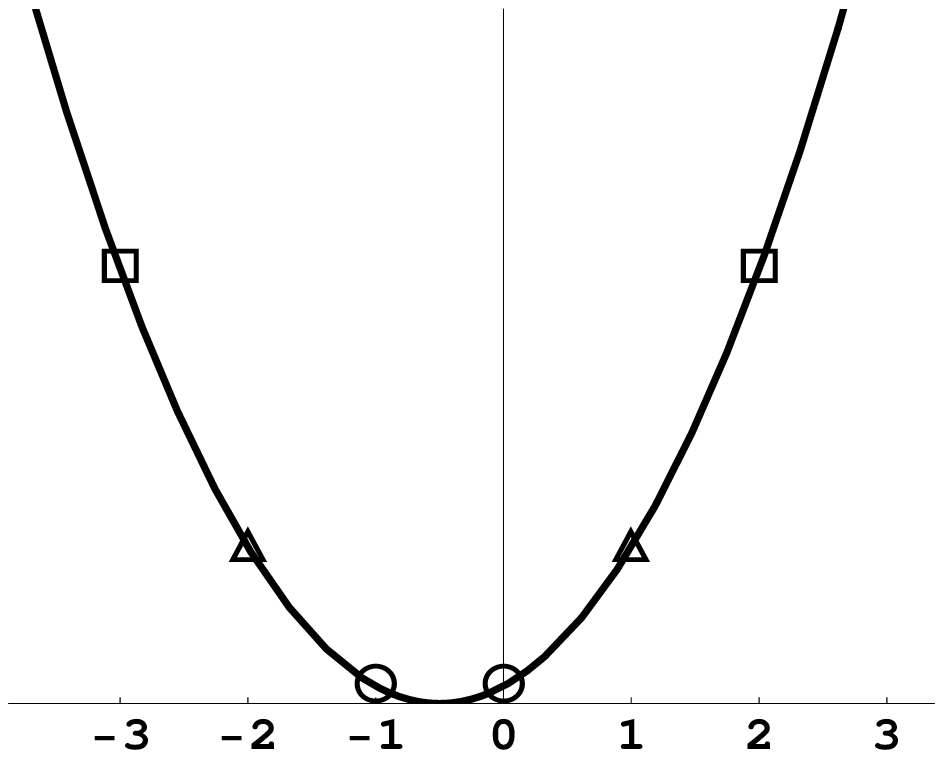}
} }
 \caption{Pairing of states~\cite{Schrieffer}. (a) Upper panel: the external flux is $\Phi=0$; the pairing is between $n_1+n_2=0$.
 (b) Lower panel: $\Phi=h/2e$; the pairing is between $n_1+n_2=-1$. The pairing configuration changes
 from type (a) to type (b) at external flux $\Phi=h/4e$; see Ref.~\cite{Schrieffer}. } \label{fig:pairing}
\end{figure}
We can expand $G$ and $F$ in Fourier series, as follows:
\begin{subequations}
\label{eqn:GnF}
\begin{eqnarray}
&&G(x_1,x_2;\omega_n)=\frac{1}{L}\sum_{n_1,n_2} g_{n_1,n_2}( \omega_n)
e^{i\frac{2\pi}{L} n_1 x_1 +i\frac{2\pi}{L} n_2 x_2},\\
&&F^\dagger(x_1,x_2;\omega_n)=\frac{1}{L}\sum_{n_1,n_2} f^\dagger_{n_1,n_2}(
\omega_n) e^{i\frac{2\pi}{L} n_1 x_1 +i\frac{2\pi}{L} n_2 x_2},
\end{eqnarray}
\end{subequations}
where $n_1$ and $n_2$ are integers labeling single-particle states. Due to the
translational (to be more precise, rotational) invariance of the system, we
assume that $G(x,x)$ has no dependence on $x$. This sets constraints on the
nonzero Fourier components for $G$: $n_1+n_2=0$. Furthermore, we assume that
$\Delta(x)= e^{i 2\pi m x/L} \Delta_0$, and hence $F^\dagger(x,x)\sim e^{-i
2\pi m x/L}\Delta_0^*$. This sets constraints on the nonzero Fourier
components for $F^\dagger$: $n_1+n_2=-m$. The meaning of this is that the
pairing occurs between the single-particle states  $n_1$ and
$n_2=-m-n_1$~\cite{Schrieffer}; see Fig.~\ref{fig:pairing}.

The Gor'kov equations can be expressed in terms of the Fourier components of
$G$ and $F^\dagger$ as follows:
\begin{subequations}
\begin{eqnarray}
&&\Big[+i\hbar\omega_n-\frac{\hbar^2}{2MR^2}\big(n_1+m-\phi\big)^2+\mu\Big]g_{n_1+m,-n_1-m}+
\Delta_0\,
f^\dagger_{n_1,-n_1-m}= \hbar, \\
&&\Big[-i\hbar\omega_n-\frac{\hbar^2}{2MR^2}\big(n_1+\phi\big)^2+\mu\Big]f^\dagger_{n_1,-n_1-m}-\Delta^*_0\,\, g_{n_1+m,-n_1-m}= 0,
\end{eqnarray}
\end{subequations}
where $\phi\equiv \Phi/(-hc/e)=\Phi/(hc/|e|)$. In the following we shall set
$\hbar=1$ , $c=1$, and $k_B=1$, for the sake of convenience. But we shall
refer to the single-particle flux quantum $hc/e$ as $h/e$~\cite{footnotehe}.
These equations can be solved explicitly, yielding
\begin{subequations}
\begin{eqnarray}
\!\!\!\!\!\!\!\!\!g_{n_1+m,-n_1-m}&=&
\frac{-i\omega_n-\myOmega(n_1\!+\!\phi)^2+\mu}{\big[i\omega_n-\myOmega(n_1\!+\!m\!-\!\phi)^2+\mu\big]\big[-i\omega_n-\myOmega(n_1\!+\!\phi)^2+\mu\big]+|\Delta_0|^2}, \\
\!\!\!\!\!\!\!\!\!f^\dagger_{n_1,-n_1-m}&=&
\frac{\Delta^*_0}{\big[i\omega_n-\myOmega(n_1\!+\!m\!-\!\phi)^2+\mu\big]\big[-i\omega_n-\myOmega(n_1\!+\!\phi)^2+\mu\big]+|\Delta_0|^2},
\end{eqnarray}
\end{subequations}
where, for the sake of convenience, we have introduced $\myOmega\equiv
1/2MR^2$ (noting that we have set $\hbar=1$ in $\hbar^2/2MR^2$). The
self-consistency equation~(\ref{eqn:selfconsistency0}) then becomes
\begin{equation}
\Delta^*_0=\frac{VT}{L}\sum_{\omega_n}
\sum_{n_1}f^\dagger_{n_1,-n_1-m}(\omega_n).
\end{equation}

 We  mainly discuss the
case in which the chemical potential $\mu$ is kept fixed, e.g.~by contact with
a particle reservoir, or else we assume that the variation of $\mu$ with
temperature and flux is sufficiently weak to be negligible near the
superconducting transition.

\subsection{Critical temperature}
To solve for $T_c(\phi)$ we set $\Delta_0=0$ in the self-consistency equation, thus obtaining
\begin{equation}
\label{eqn:tc}
 1=\frac{VT}{L}\sum_{\omega_n}\sum_{n_1\in
 Z}\frac{1}{[i\omega_n-\myOmega(n_1\!+\!m\!-\!\phi)^2+\mu][-i\omega_n-\myOmega(n_1\!+\!\phi)^2+\mu]}.
\end{equation}
It is important to note the underlying assumption that the transition from
superconducting to normal is associated with a vanishing order parameter, and
hence is a continuous transition. The consistency of this assumption needs to
be checked once we obtain the solution. We shall see, below, that for
sufficiently small radii the assumption is not valid and the transition is
actually associated with non-vanishing order parameter, and hence is
discontinuous.

 Next, we make use of the root of the
Poisson summation formula (i.e.~the Dirac comb and its Fourier series),
$\sum_{n_1\in Z}\delta(x-n_1)=\sum_{k \in Z} e^{i2\pi x k}$, to turn the
summation over $n_1$ in Eq.~(\ref{eqn:tc}) into one over the conjugate
variable $k$:
\begin{eqnarray}
\label{eqn:PoissonSum}
&&\sum_{n_1}\frac{1}{[i\omega_n-\myOmega(n_1\!+\!m\!-\!\phi)^2+\mu][-i\omega_n-\myOmega(n_1\!+\!\phi)^2+\mu]}\cr
=&&\sum_{k \in Z} \int_{-\inf}^{\inf} dx \,\,\frac{e^{i2\pi x
k}}{[i\omega_n-\myOmega(x\!+\!m\!-\!\phi)^2+\mu][-i\omega_n-\myOmega(x\!+\!\phi)^2+\mu]}\,\,\,.
\end{eqnarray}
Instead of placing a cutoff on the energy, we follow Gor'kov~\cite{Gorkov} and
place the cutoff (which, in practice, is the Debye frequency $\omega_D$) on
the Matsubara frequency.

\subsubsection{Large-radius limit}
\label{sec:LargeRadiusClean}
 For sufficiently large $R$ we can
ignore the correction terms associated with finite radius (i.e.~$k\ne 0$), and
thus we obtain an equation relating the critical temperature at nonzero flux
to that at zero flux. In the limits that the Debye frequency is much smaller
than the chemical potential (i.e.~$\omega_D/\mu\ll 1$), and that the chemical
potential is much larger than the level spacing (i.e.~$\sqrt{2MR^2\mu} \gg
1$), we  obtain (see Appendix~\ref{app:cleanlarge} for the derivation)
\begin{eqnarray}
\label{eqn:TcBulk}
\ln\left(\frac{T_\text{c}(\phi)}{T^0_\text{c}}\right)=\psi\left(\frac{1}{2}\right)-{\rm
Re}\,\,\psi\left(\frac{1}{2}-i \frac{x_m(\phi)}{2\pi
T_\text{c}}\sqrt{\frac{2\mu}{MR^2}} \right),
\end{eqnarray}
where  $T^0_\text{c}\equiv T_\text{c}(0)$ is the critical temperature at zero
flux for the same radius, $x_m(\phi)\equiv \phi-m/2$ [with $m$ being chosen to
minimize $(2\phi-m)^2$, i.e.~the kinetic energy in the Ginzburg-Landau
picture], and $\psi(x)$ is the digamma function,
\begin{equation}
\label{eqn:digamma}
 \psi(x)\equiv-\gamma +\sum_{k=0}^\infty\big(
\frac{1}{k+1}-\frac{1}{k+x}\big),
\end{equation}
where $\gamma$ is the Euler constant~\cite{AbramowitzStegun}.

 The zero-flux, zero-temperature order parameter $\Delta_0$ is related to
the corresponding critical temperature $T^0_\text{c}$ in the weak-coupling
limit by $\Delta_0=\Gamma\, T^0_\text{c}$, where $\Gamma=\pi/e^\gamma\approx
1.76$. We call the length $\xi_0\equiv v_F/\pi\Delta_0$ the Cooper-pair size
(where $v_F$ is the Fermi velocity), so as to distinguish it from the
zero-temperature Ginzburg-Landau coherence length $\xi(0)$. There are two
further relevant energy scales: the chemical potential $\mu$, and the
single-particle energy-level spacing $\myOmega$. The ratio
${\myOmega\mu}/{\Delta_0^2}$ can be estimated as
\begin{equation}
\label{eqn:ratio}
 \frac{\myOmega\,\mu}{\Delta_0^2}\approx
\left(\frac{\hbar^2}{2MR^2}\right)\left(\frac{M v_F^2}{2}\right)\Big/
\left(\frac{\hbar v_F}{\pi\xi_0}\right)^2=\left(\frac{\pi\xi_0}{2R}\right)^2 ,
\end{equation}
which is set by on the ratio of the ring radius to the Cooper-pair size. This
motivates us to define a measure $\rho$ of the ratio of the radius to the
Cooper-pair size, via
\begin{equation}
\label{eqn:rho}
 \left(\frac{\pi}{2\rho}\right)^2\equiv
\frac{\myOmega\,\mu}{\Delta_0^2}.
\end{equation}
Then, defining $t\equiv T_\text{c}/T^0_\text{c}$, we can re-write the equation
for the critical temperature as
\begin{eqnarray}
\label{eqn:TcClean2} \ln t=\psi\left(\frac{1}{2}\right)-{\rm
Re}\,\psi\left(\frac{1}{2}-i \frac{x_m(\phi) \Gamma}{2\rho\, t} \right).
\end{eqnarray}
\begin{figure}

\vspace{0.5cm} \centerline{ \rotatebox{0}{
        \epsfxsize=7.0cm
        \epsfbox{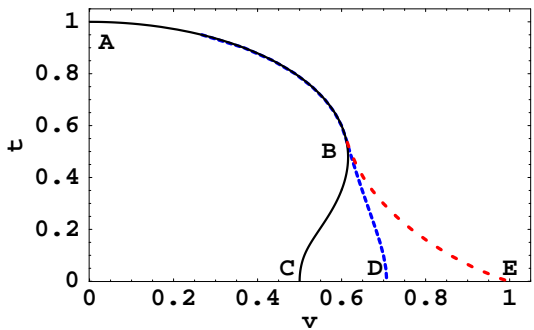}
} }
 \caption{Phase diagram: reduced critical temperature $t(\phi,R)=T_c(\phi,R)/T_c(0,R)$ vs.~flux and radius
 $v=\pi |\phi-m/2|(\xi_0/R)$ in the clean limit and in the limit that the finiteness corrections are
 ignored (i.e.~bulk limit). In the case of the exchange-field effect, discussed in Refs.~\cite{Sarma,MakiTsuneto},
 the horizontal axis becomes $v=\mu_B H/\Delta_0$, where $\mu_B$ is Bohr magneton and $H$ is the exchange field.
 The curve ABC is the solution to Eq.~(\ref{eqn:TcClean2}); ABD is the curve representing the true
 equilibrium transition line. From A to B the transition is continuous, whereas
 from B to D the transition is
 discontinuous, a construction first made in Refs.~\cite{Sarma,MakiTsuneto} in the context of the
 exchange-field effect. The curve BE represents the metastability limit for ``superheating'', whereas
 the curve BC represents the metastability limit for ``supercooling''~\cite{MakiTsuneto}. } \label{fig:sfPhase}
\end{figure}
An equation of this form was studied by Sarma~\cite{Sarma} and by Maki and
Tsuneto~\cite{MakiTsuneto}, both in the context of the effect of a magnetic
exchange field on superconductivity. The correspondence is that the role of
the exchange field normalized to the zero-temperature gap (i.e.~$\mu
H/\pi\Delta_0$) is, in the present setting, played by the combination of the
normalized inverse ring radius and the flux (i.e.~$x_m(\phi)/\rho$). In
Refs.~\cite{Sarma,MakiTsuneto} it was found that for large enough exchange
field (here, small enough ring radius) Eq.~(\ref{eqn:TcClean2}) has multiple
solutions and that, moreover, the correct interpretation is that the
transition between the normal and superconducting states becomes discontinuous
(beyond certain value of the exchange field). By contrast, for small enough
exchange field (here, large enough ring radius) the transition is continuous.
By borrowing the results of Refs.~\cite{Sarma,MakiTsuneto}, as summarized in
Fig.~\ref{fig:sfPhase}, we have that the threshold at which the transition
changes character between continuous and discontinuous occurs at
$|x_m(\phi)|/\rho \approx 0.6/\pi$ (i.e.~point B in Fig.~\ref{fig:sfPhase}).
Furthermore, one has that for extremely small rings, such that
$|x_m(\phi)|/\rho > 1/\sqrt{2}\pi$ (i.e.~point D), the system never becomes
superconducting~\cite{footnoteSarma}. Of course, these estimates would need to
be modified if the finiteness of the radius were to be taken into account, but
we expect that the qualitative separation into discontinuous- and
continuous-transition regimes would still hold.
\begin{figure}

\vspace{0.5cm} \centerline{ \rotatebox{0}{
        \epsfxsize=7.0cm
        \epsfbox{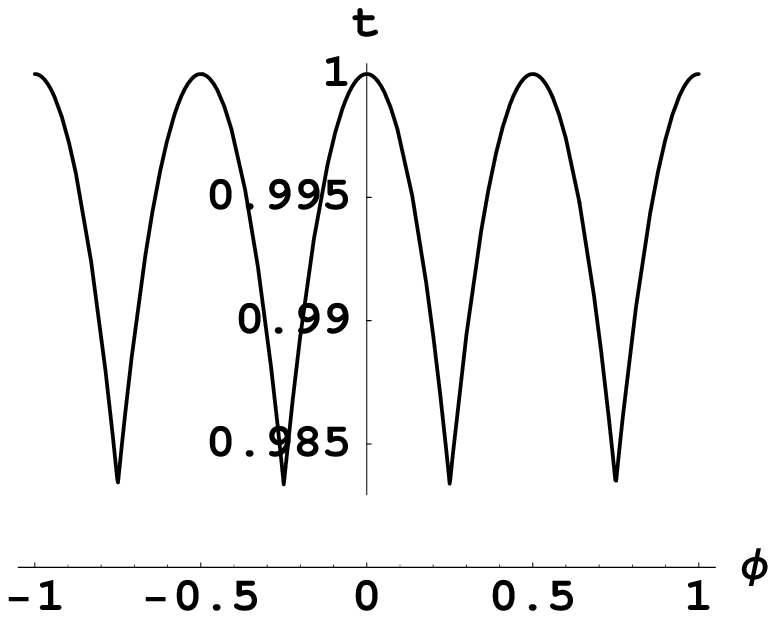}
} } \centerline{ \rotatebox{0}{
        \epsfxsize=7.0cm
        \epsfbox{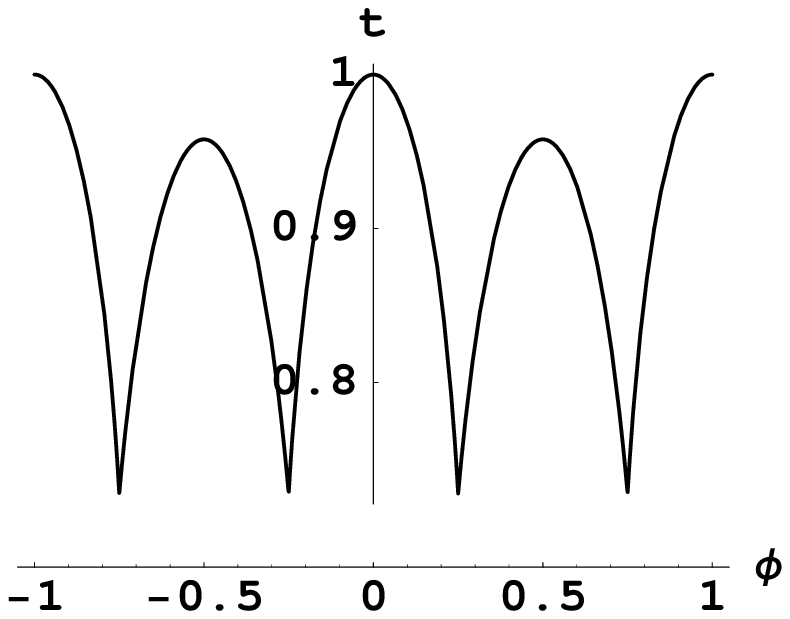}
} } \vspace{-0.2cm} \caption{Critical temperature (normalized to its zero-flux
value) $t$ vs.~flux $\phi=\Phi/(h/e)$ for $R=5\,\xi_0$ (upper) and
$R=1.5\,\xi_0$ (lower) in the clean limit. The plots are made with
$\cos\big(2\pi
 \overline{n}\big) =1$ and with only the $k=1$ term in Eq.~(\ref{eqn:Tcclean}) retained. } \label{fig:1}
\end{figure}

 For large-radius rings (i.e.~$\rho\gg 1$), we can
expand the second digamma function in Eq.~(\ref{eqn:TcClean2}) to second order
and the logarithm to first order, thus obtaining
\begin{eqnarray}
-(1-t)\approx \frac{1}{2}\psi''\left(\frac{1}{2}\right)\left(\frac{x_m(\phi)
\Gamma}{2\rho\, t}\right)^2\approx -8.41 \left(\frac{x_m(\phi) \Gamma}{2\rho\,
t}\right)^2.
\end{eqnarray}
Hence, we see that the fractional reduction in the critical temperature due to
the flux is given by
\begin{eqnarray}
\label{eqn:expand} (1-t)\approx 8.41 \left(\frac{x_m(\phi) \Gamma}{2\rho\,
t}\right)^2\approx 8.41 \left(\frac{x_m(\phi) \Gamma}{2\rho}\right)^2= 8.41
\left(\frac{ \Gamma}{4\rho}\right)^2(2\phi-m)^2.
\end{eqnarray}
The integer $m$ must to be chosen such that $(2\phi-m)^2$ is minimum, so as to
obtain the most stable solution; hence, we recover the standard Little-Parks
oscillation result, for which the period  is $h/2e$.
Equation~(\ref{eqn:expand}) can be re-expressed as
\begin{equation}
\label{eqn:min} \sqrt{\frac{1-t}{8.41}}\frac{4R}{\Gamma\xi_0}=\min_{m\in
Z}|2\phi-m|.
\end{equation}
The l.h.s.~can be much smaller than unity if $R$ is much smaller than $\xi_0$,
but the value of the r.h.s.~can range from $0$ to $1/2$, depending on the
value of external flux. This means that there can exist a range of fluxes for
which no solution of $t$ exists. This reflects the fact that superconductivity
is destroyed over certain ranges of flux~\cite{deGennes}. To make connection
with the result by de Gennes~\cite{deGennes}, let us multiply
Eq.~(\ref{eqn:min}) by $2\pi$ and take the cosine of both sides. We recover
the de Gennes result for the transition temperature (for the case in which the
length of the side arm in Ref.~\cite{deGennes} is set to zero)
\begin{equation}
\label{eqn:dGclean}
 \cos\left(2\pi\frac{R}{\xi(t)}\right)\approx
\cos\left(2\pi\frac{\Phi}{h/2e}\right),
\end{equation}
where $\xi(t)\approx 0.74\Gamma\, \xi_0/\sqrt{1-t}$ is the
temperature-dependent Ginzburg-Landau coherence length in the clean limit.
 We note that, however, the
Ginzburg-Landau approach is, strictly speaking, valid only near $t\approx 1$.
Furthermore, the existence of discontinuous transitions is beyond the reach of
the Ginzburg-Landau approach.

\subsubsection{ Finite-radius correction}
\begin{figure}[h]

\vspace{0.5cm} \centerline{ \rotatebox{0}{
        \epsfxsize=7.0cm
        \epsfbox{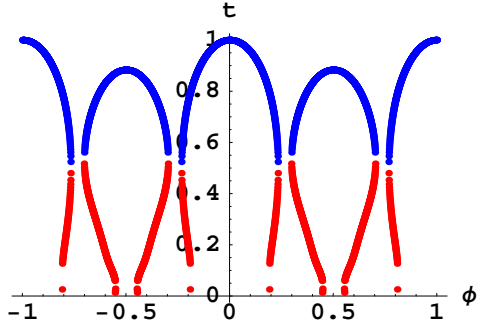}
} } \caption{Reduced critical temperature $t(\phi)\equiv
T_\text{c}(\phi)/T_\text{c}(0)$ (normalized to its zero-flux value) vs.~flux
$\phi\equiv\Phi/(h/e)$ for $R=1.25\, \xi_0$  with $\cos\big(2 k\pi
 \overline{n}\big) =1$ in the clean limit. The higher $T_c$ branches (blue) are calculated by retaining only the first
 two terms in Eq.~(\ref{eqn:Tcclean}), whereas the lower $T_c$ branches (red) are calculated by retaining as many as ten such terms.
 The plot shows multiple solutions to Eq.~(\ref{eqn:Tcclean}),
 and signifies the emergence of discontinuous transitions. The upper and lower branches of
 the solutions are expected to merge at certain values of $\phi$; the appearance of a gap between them is
 an artifact of our considering only a  finite number of values of $\phi$.} \label{fig:multiple}
\end{figure}
\label{sec:clean_finite}  What is the correction to $T_{\text{c}}$ that arises
from the finiteness of the radius? By taking into account this correction we
arrive at the following equation obeyed by the critical temperature  (see
Appendix~\ref{app:cleanfinite} for the derivation):
\begin{eqnarray}
\label{eqn:Tcclean}
 \ln t(\phi)&=&\psi\left(\frac{1}{2}\right)\!-\!{\rm
Re}\,\psi\left(\frac{1}{2}\! -\! \frac{i x_m(\phi) \Gamma}{2\rho\, t(\phi)}
\right) - 4\sum_{k=1}^\infty  \left\{ \cos(2\pi k
 \overline{n}) \,e^{-\frac{2\pi k \rho }{\Gamma}} f_c(0,\rho,1) \right. \nonumber \\
\medskip &&\quad\qquad\left. -  {\rm Re}\Big[e^{i2\pi
k\overline{n}} e^{-\frac{2\pi k\rho
t}{\Gamma}}f_c\big(x_m(\phi),\rho,t(\phi)\big) \cos2\pi k\phi\Big]\right\},
\end{eqnarray}
where we recall that $t(\phi)=T_\text{c}(\phi)/T_\text{c}(0)$, $\Gamma$ is the
BCS constant $\Gamma=\pi/e^{\gamma}\approx 1.76$, we have defined
$\overline{n}\equiv \sqrt{2MR^2\mu} $, which is related to the number of
Cooper pairs, and the function $f_c$ is defined via the hypergeometric
function ${}_2F_1(a, b; c; z)$~\cite{AbramowitzStegun}:
\begin{equation}
f_c\big(x_m(\phi),\rho,t\big)\equiv{}_2F_1\Big(\frac{1}{2}\!-\! \frac{i
x_m(\phi) \Gamma}{2\rho\, t},1,\frac{3}{2}\!-\! \frac{i x_m(\phi)
\Gamma}{2\rho\, t},e^{-\frac{4\pi k\rho t}{\Gamma}}\Big)\big/\big(1\!-\!
\frac{i x_m(\phi) \Gamma}{\rho\, t}\big)
\end{equation}
In Figs.~\ref{fig:1} and~\ref{fig:multiple} we show the flux dependence of the
critical temperature for two particular ring radii. For the larger radius
case, the value of $t$ at $\Phi=h/2e$ is essentially the same as that at
$\Phi=h/e$ and $0$;  for the smaller radius case and for
$\cos(2\pi\overline{n})=1$ (i.e.~all the pair states are occupied at and below
Fermi level), the amplitude at $\Phi=h/2e$ is reduced, relative to that at
$\Phi=h/e$ and $0$. Thus, as the radius becomes small we clearly see the
emergence of the single-particle flux quantum period $h/e$. Also worth
noticing is the occurrence of a second solution (with lower value) for the
critical temperature at sufficiently small radii, shown in
Fig.~\ref{fig:multiple}. Compared to the higher $T_c$ solutions, for which our
approximation of the $T_c$ equation~(\ref{eqn:Tcclean}), by keeping only the
first two correction terms yields rather precise values, the evaluation of
$T_c$ at these lower values of temperatures requires more terms (e.g.~four or
more) to be included in order to maintain the same accuracy.

\begin{figure}

\vspace{0.5cm} \centerline{ \rotatebox{0}{
        \epsfxsize=7.0cm
        \epsfbox{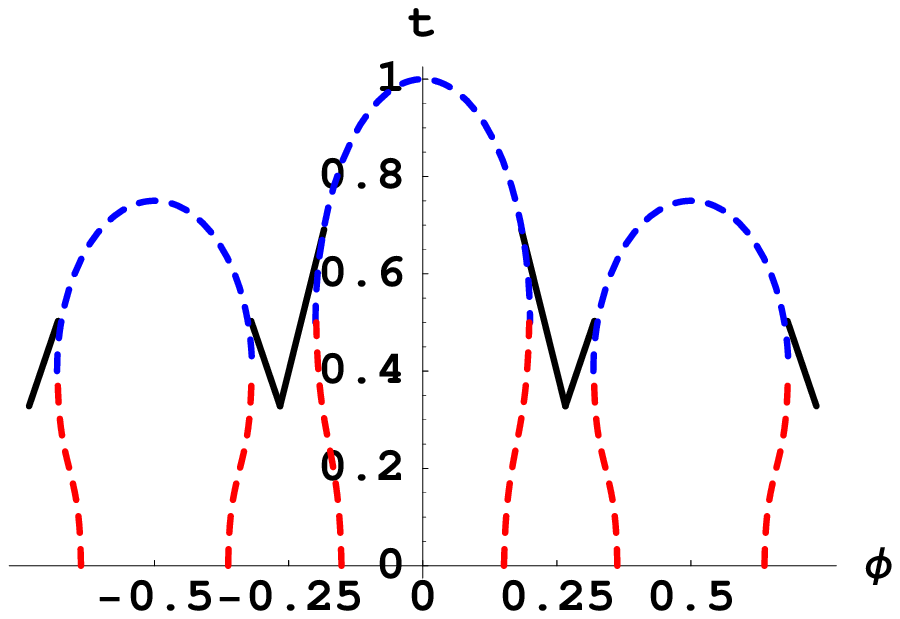}
} }
 \centerline{ \rotatebox{0}{
        \epsfxsize=7.0cm
        \epsfbox{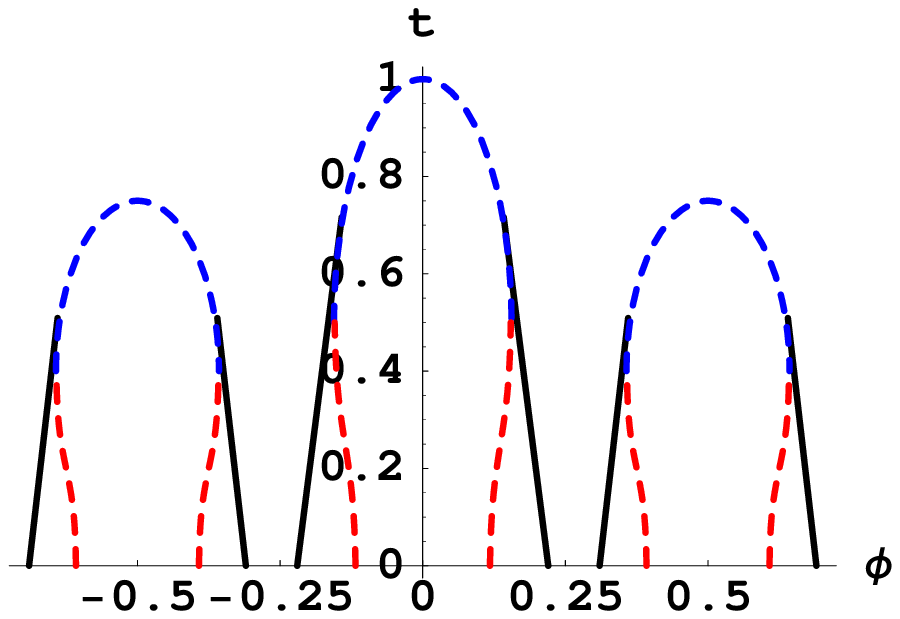}
} } \caption{Schematic depiction of the reduced critical temperature
$t(\phi)\equiv T_\text{c}(\phi)/T_\text{c}(0)$ (normalized to its zero-flux
value) vs.~flux $\phi=\Phi/(h/e)$ for small radius rings in the clean limit.
As illustrated in Fig.~\ref{fig:multiple}, for sufficiently small radii, there
are multiple solutions for $T_\text{c}$, as exemplified there by the higher
(blue) branches and the lower (red) branches. Near $\phi=0$ and $\pm1/2$, the
upper branches are the equilibrium phase boundary and the transition to normal
state is continuous. Away from these regions, the globally stable equilibrium
phases must be sought by free-energy consideration, and the correponding phase
boundaries are indicated schematically by the solid lines and represent
discontinuous transitions. Upper panel: For all flux values, there exists a
superconducting state. Lower panel: For smaller radius, it can happen that
there are flux values for which no superconducting state exists.}
\label{fig:schematic}
\end{figure}
For the case of large rings (i.e.~$\rho\gg 1$), to determine the leading
corrections it is adequate to  retain only the $k=1$ term and to set the
correction $\sim i x_m(\phi) \Gamma/\rho t$ to zero, when compared to values
of order unity, in the second $f_c$ function. Thus, we arrive at the formula
\begin{eqnarray}
\ln t(\phi)\approx&\psi\left(\frac{1}{2}\right)-{\rm
Re}\,\psi\left(\frac{1}{2} -i \frac{x_m(\phi) \Gamma}{2\rho\, t}
\right)-4\cos(2\pi\overline{n})\,
e^{-2\pi\rho/\Gamma}f(0,\rho,1)[1-\cos2\pi\phi].
\end{eqnarray}
The change in the reduced critical temperature $t(\phi)$  is then
approximately given by
\begin{equation}
1-t(\phi)\approx 8.41 \left(\frac{ \Gamma}{4\rho}\right)^2(2\phi-m)^2+4
\cos(2\pi
 \overline{n}) \tanh^{-1}\big(e^{-\frac{2\pi \rho }{\Gamma}}\big) (1-\cos2\pi
\phi),
\end{equation}
where we have used the fact that ${}_2F_1(1/2,1,3/2,y^2)=
\tanh^{-1}({y})/{y}$. Again, the integer $m$ is to be chosen to minimize
$(2\phi-m)^2$. We see that, in addition to the parabolic dependence on $\phi$,
there is a sinusoidal correction of period $h/e$. This is the emergence of the
single-particle flux dependence. We also note that this correction is not
universal, in that it depends sensitively on the value of $\mu$ [and,
moreover, the form of $\cos\big(2\pi\overline{n})$ results from the simple
quadratic single-particle spectrum]. It can happen that the correction due to
the finiteness of the radius
 actually increases the critical temperature, i.e.~when
$\cos\big(2\pi\overline{n})<0$.

As we have argued using the results of Sarma~\cite{Sarma} and Maki and
Tsuneto~\cite{MakiTsuneto}, the occurrence of multiple solutions in
Eq.~(\ref{eqn:TcClean2}) for $T_\text{c}(\phi)$, for certain ranges of
$|x_m(\phi)|/\rho$, leads to a change from a continuous to a discontinuous
superconducting-to-normal phase transition. Even with the corrections to
$T_{\text{c}}$ due to the finite-radius effect, as the radius of the ring
decreases (to a value comparable to the coherence length), we observe that
Eq.~(\ref{eqn:Tcclean}) still possesses multiple solutions for
$T_\text{c}(\phi)$ for certain ranges of $|x_m(\phi)|/\rho$ ;  see
Fig.~\ref{fig:multiple}. This implies that somewhere in these ranges (of
$|x_m(\phi)|/\rho$) there exists a change from continuous (at larger radius)
to discontinuous (at smaller radius) superconducting-to-normal transition.
This is shown schematically in Fig.~\ref{fig:schematic}. If the radius is
sufficiently large, the transition is always continuous. If the radius is
sufficiently small, the the curve representing the discontinuous transition in
one ``dome'' can intersect with that of the nearby dome (above the ``void''
region where no solution for $T_\text{c}$ of Eq.~(\ref{eqn:Tcclean}) exists).
If this void region is large, the curve of the discontinuous transition can go
to $t=0$ at certain value of $\phi$ without intersecting that from the nearby
dome. However, calculations of $T_c(\phi)$ alone cannot determine the precise
location of the change from continuous to discontinuous; considerations of
free energies are necessary to settle this issue.

\section{Disordered regime}
\label{sec:disordered}
\begin{figure}
\vspace{0.5cm} \centerline{ \rotatebox{0}{
        \epsfxsize=7.0cm
        \epsfbox{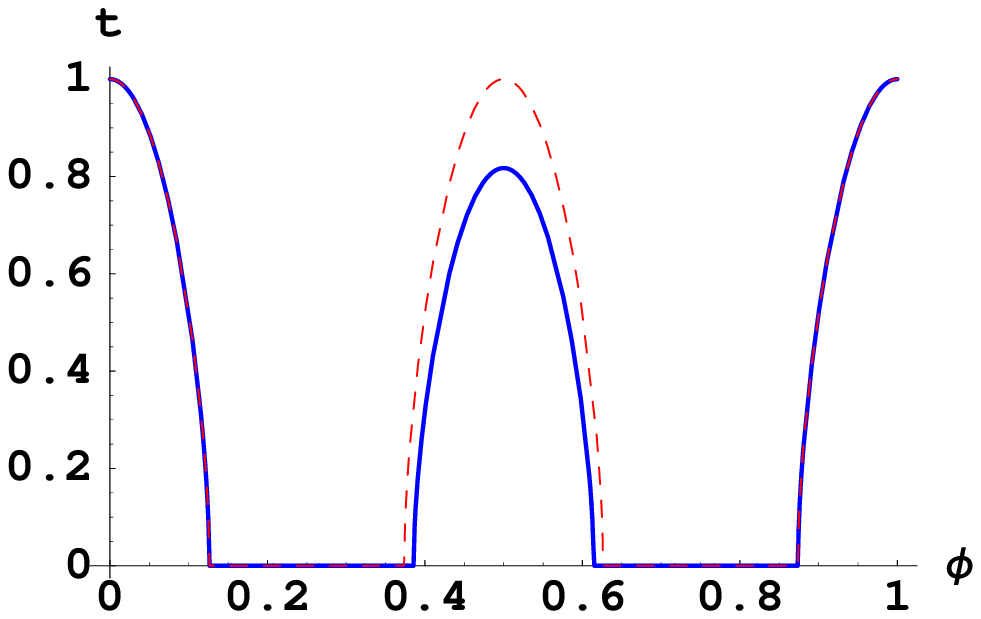}
} }
 \centerline{ \rotatebox{0}{
        \epsfxsize=7.0cm
        \epsfbox{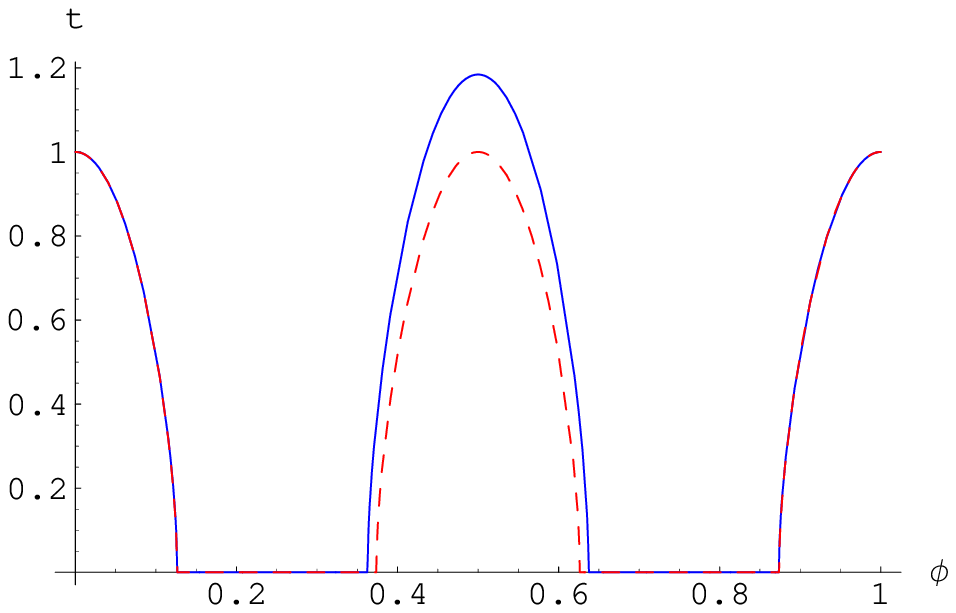}
} }
 \caption{Critical temperature (normalized to its zero-flux value) for the disordered regime vs.~flux $\phi=\Phi/(h/e)$ for $R=0.2\xi_0$ and the mean-free path $l_e=0.2 \xi_0$
(solid blue line) and for the
 same radius but $ l_e\ll R$ (dashed red line). The plots are made with
 $\cos\big(2\pi k
 \overline{n}\big)=1$ and $(-1)^k$ for the upper and lower panels, respectively.
 Moreover, only
 the  $k=0, 1$ and $2$ terms in Eq.~(\ref{eqn:finite_dirty}) have been retained, as the convergence is rather good.  } \label{fig:dirty}
\end{figure}
How does disorder, in the form of potential scattering from fixed impurities,
alter the physical picture that we have obtained so far? Does it affect the
critical temperature? Does it change the order of the
superconducting-to-normal transition? We now address these issues, assuming
that the configuration of the impurities is sufficiently random  that the
tendency towards CDW formation is not enhanced.
\subsection{Expansion of anomalous Green function; disorder average}
For the most part, we shall consider impurities that produce scalar potential
scattering; {\it mutatis mutandis\/}, the effects of exchange and spin-orbit
scattering can be straightforwardly included. To begin with, we include
potential scattering via the $V(x)$ term in the Gor'kov
equations~\cite{Gorkov}, which now read (here we restore the constants $\hbar$
and $c$)
\begin{subequations}
\begin{eqnarray}
&&\!\!\!\!\!\!\!\!\!\!\!\!\!\!\!\!\!\!\Big[+i\hbar\omega_n-\frac{1}{2M}\big(i\hbar\partial_x+\frac{e
\Phi}{c
L}\big)^2-V(x)+\mu\Big]G(x,x';\omega_n)+\Delta({x})F^\dagger(x,x';\omega_n)=
\hbar\delta(x-x'), \\
&&\!\!\!\!\!\!\!\!\!\!\!\!\!\!\!\!\!\!\Big[-i\hbar\omega_n-\frac{1}{2M}\big(i\hbar\partial_x-\frac{e \Phi}{c
L}\big)^2-V(x)+\mu\Big]F^\dagger(x,x';\omega_n)-\Delta^*(x)G(x,x';\omega_n)=
0,
\end{eqnarray}
\end{subequations}
where $V(x)$ represents the potential from static impurities, i.e.,
\begin{equation}
V(x)=\sum_a u(x-x_a),
\end{equation}
and $x_a$ indicates the spatial location of $a^{\rm th}$ impurity.

Following Gor'kov~\cite{Gorkov}, we introduce a Green function
$G^0(x,x';\omega_n)$ that satisfies
\begin{equation}
\Big[i\hbar\omega_n-\frac{1}{2M}\big(i\hbar\partial_x+\frac{e \Phi}{c
L}\big)^2-V(x)+\mu\Big]G^0(x,x';\omega_n)= \hbar\delta(x-x').
\end{equation}
We can then express $F^\dagger$ exactly in terms of $\Delta^*$ and $G^0$ as
\begin{equation}
\label{eqn:Fd}
 F^\dagger(x_1,x_2;\omega_n)=\frac{1}{\hbar}\int dx\,
G^0(x,x_1;\omega_n)\,\Delta^*(x)\,G(x,x_2;-\omega_n).
\end{equation}
Because we are only concerned with solving for $T_\text{c}$, near the
transition, it is sufficient to keep G to zeroth order in $\Delta$ and replace
$G$ in Eq.~(\ref{eqn:Fd}) by $G^0$; thus we have
\begin{equation}
F^\dagger(x_1,x_2;\omega_n)\approx\frac{1}{\hbar}\int dx\,
G^0(x,x_1;\omega_n)\,\Delta^*(x)\,G^0(x,x_2;-\omega_n).
\end{equation}
We now consider the self-consistency equation~(\ref{eqn:selfconsistency0}),
and average over the quenched disorder associated with the locations of the
impurities:
\begin{equation}
\overline{\Delta^*(r)}=\frac{V}{\beta}\sum_{\omega_n}\int dx\,
\overline{G^0(x,r;\omega_n)\,G^0(x,r;-\omega_n)}\,\overline{\Delta^*(x)},
\end{equation}
where $\overline{\cdots}$ indicates  disorder averaging. We note that, as
explained by Gor'kov~\cite{Gorkov}, the Green function $G^0$ oscillates on a
much smaller length scale than $\Delta^*$ and, hence, the disorder average of
$\Delta^*$ can be factorized.  For convenience, we
 use $\widetilde{G}$ to denote the disorder average of $G^0$, i.e.,
$\widetilde{G}(x,r;\omega_n)\equiv\overline{G^0(x,r;\omega_n)}$, the
translational invariance of which is restored, viz.,
\begin{equation}
\widetilde{G}(x,r;\omega_n)=\frac{1}{L}\sum_{n_1} \widetilde{G}(n_1;\omega_n)
e^{i\frac{2\pi n_1}{L}(x-r)}.
\end{equation}

To calculate the disorder average of the product of two Green functions, we
introduce the kernel $K$~\cite{Gorkov}, defined via
\begin{equation}
\overline{G^0(x,r;\omega_n)\,G^0(x,r';-\omega_n)}\equiv
\frac{1}{L^2}\sum_{n_1,n_2\in Z} K_{\omega_n}(n_1,n_2) e^{i\frac{2\pi
n_1}{L}(x-r)}e^{i\frac{2\pi n_2}{L}(x-r')}.
\end{equation}
If we retain only the ladder diagrams (i.e.~ignoring the crossed
diagrams~\cite{AG59}), we arrive at the result
\begin{equation}
\label{eqn:Kernel} K_\omega(n_1,n_2)=\widetilde{G}_\omega(n_1)
\,\widetilde{G}_{-\omega}(n_2)\big[1+ n_{\rm imp}\sum_q |u(q)|^2
K_\omega(n_1-q,n_2+q)\big],
\end{equation}
where we have simplified the notation by dropping the subscript $n$ on
$\omega$ and moving $\omega$ from an argument to a subscript, and $n_{\rm imp}$ is the impurity concentration. Assuming that the
order parameter retains the form $\overline{\Delta^*(r)}=\Delta^*_0\,
e^{-i2\pi m r/L}$, the self-consistency equation can be reduced to
\begin{equation}
\label{eqn:Self-consistency1}
1=\frac{V}{\beta L}\sum_\omega \sum_{n_1,n_2} \delta_{n_1+n_2,m} K_\omega(n_1,n_2).
\end{equation}

\begin{figure}\vspace{0.5cm} \centerline{ \rotatebox{0}{
        \epsfxsize=7.0cm
        \epsfbox{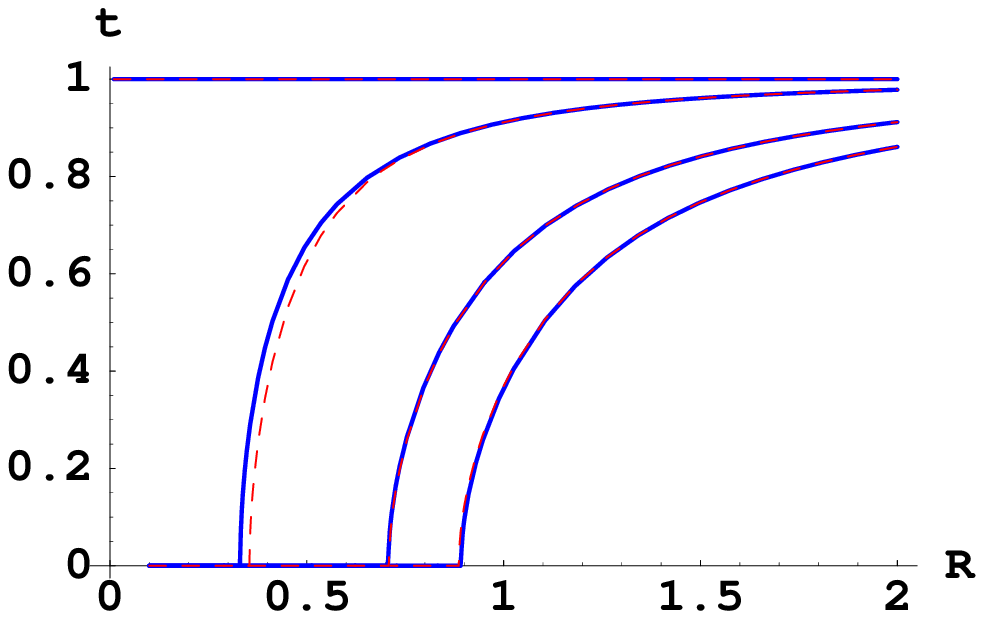}
} }
\vspace{0.5cm}
 \centerline{ \rotatebox{0}{
        \epsfxsize=7.0cm
        \epsfbox{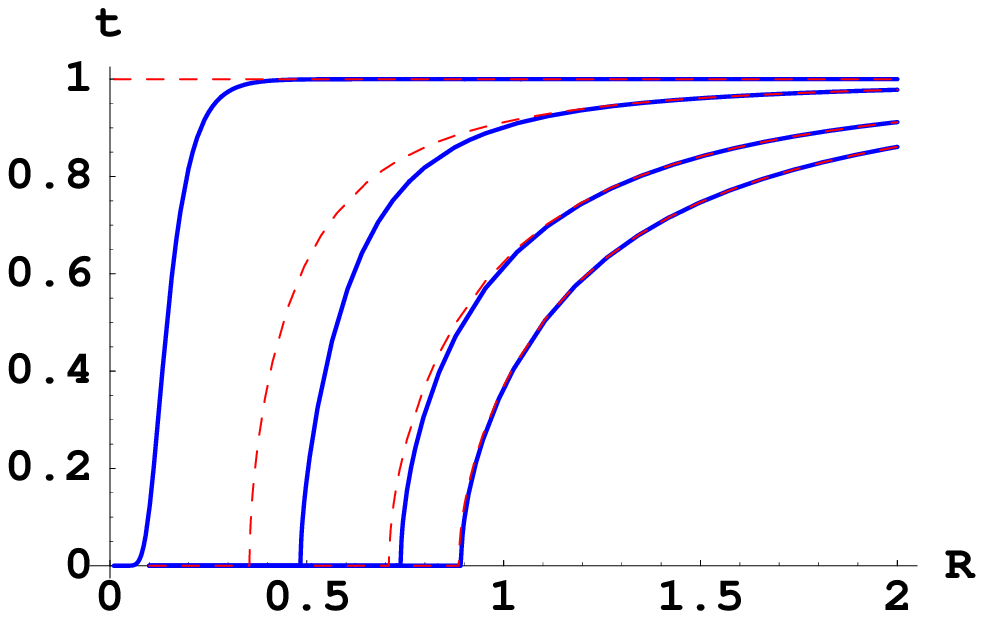}
} }
 \caption{Phase boundaries between normal state (to the left of the curves) and the superconducting state in the
 temperature ($t(\phi,R)=T_c(\phi,R)/T_c(\phi=0,R)$, normalized to its zero-flux value) vs.~radius ($R$, in unit of $\xi_0$) plane. Upper panel:~$\phi=\Phi/(h/e)=0,0.1,0.2,0.25$ (from the top
down). The solid blue boundaries take into account of the finite-radius
corrections, whereas the dashed red boundaries are solutions to the bulk
equation, Eq.~(\ref{eqn:bulk_dirty}). Lower
panel:~$\phi=\Phi/(h/e)=0.5,0.40,0.30,0.25$ (from the top down). In all plots,
we have chosen $l_e=0.2\xi_0$ and $\cos\big(2\pi k
 \overline{n}\big)=1$, and have kept terms up to $k=2$ in Eq.~(\ref{eqn:finite_dirty}).
 Note the significant deviations associated with the finite-radius corrections near $\phi=0.5$.} \label{fig:Phasedirty}
\end{figure}
\subsection{Critical temperature}
We  leave the detailed calculation of the kernel to
Appendix~\ref{app:dirtylimit} and simply quote here the resulting
equation~(\ref{app:eqn:finite_dirty}) for the critical temperature in the
disordered regime, i.e.~$\tau_0\,T_\text{c}\ll 1$:
\begin{eqnarray}
 \label{eqn:finite_dirty}
 \ln t&=&\psi\Big(\frac{1}{2}\Big)-\psi\Big(\frac{1}{2}+\frac{\Gamma
 l_e\xi_0x_m^2(\phi)}{t R^2}\Big)
 +\sum_{k=1}^\infty4 e^{-\pi k  \frac{R}{l_e}}\cos(2\pi k \overline{n} )\times
 \nonumber\\
 &&\quad\qquad\left[ e^{-\frac{2\pi k }{\Gamma}\frac{R}{\xi_0} t }\cos(2\pi k \phi)
 f_d\big(x_m(\phi),R,t(\phi)\big)-e^{-\frac{2\pi k }{\Gamma}\frac{R}{\xi_0}}f_d(0,R,1)\right],
\end{eqnarray}
where $l_e\equiv v_F \tau_0$ is the elastic mean free path, $\tau_0$ is the
elastic scattering time, we recall that $\Gamma=\pi/e^\gamma$, and
\begin{equation}
 f_d\big(x_m(\phi),R ,t(\phi)\big)\equiv {}_2F_1\Big[\frac{1}{2}+\frac{\Gamma
l_e\xi_0x_m^2}{t R^2},1,\frac{3}{2}+
 \frac{\Gamma l_e\xi_0 x_m^2}{t R^2},e^{-\frac{4\pi k
 }{\Gamma}\frac{R}{\xi_0}t}\Big]
\Big/\big(1+\frac{2\Gamma l_e\xi_0 x_m^2}{t R^2}\big).
\end{equation}
We remark that the argument in the second digamma function in
Eq.~(\ref{eqn:finite_dirty}) is real, in contrast with the clean case, for
which it is complex; see Eq.~(\ref{eqn:TcBulk}). A consequence of this is that
there is no longer a doublet of solutions for $T_\text{c}$. Moreover, the
resulting single solution is consistent with the assumption that the order
parameter becomes vanishingly small as the temperature approaches its critical
value from the superconducting side. Therefore, the transition to the
superconducting state is continuous.

To explore the consequences of Eq.~(\ref{eqn:finite_dirty}), we  first examine
the large-radius (i.e.~bulk) limit, and show that we recover the de Gennes
results for the case of rings.  We then proceed to compare how the finiteness
of the radius affects the oscillations of $T_\text{c}(\phi)$.
\subsubsection{Large-radius limit}
Ignoring correction due to the finiteness of the radius, we obtain the
following equation for $T_\text{c}$ of the bulk superconductor, i.e.,
\begin{equation}
\label{eqn:bulk_dirty} \ln\left(\frac{T_c(\phi)}{T_c^0}\right)=\ln
t(\phi)=\psi\Big(\frac{1}{2}\Big)-\psi\Big(\frac{1}{2}+\frac{\Gamma
 l_e\xi_0x_m^2(\phi)}{t(\phi) R^2}\Big),
\end{equation}
as we should. This is in agreement with the results   of de
Gennes'~\cite{deGennes} for rings and those obtained by Lopatin et
al.~\cite{LopatinShahVinokur} for hollow cylinders. To see that the de Gennes
results are recovered, we note that the critical value of $\Gamma
l_e\xi_0x_m^2(\phi)/R^2$,  beyond which no superconducting solution for any
$T>0$  exists, is given by $\Gamma/4\pi$~\cite{footnote}. For a fixed radius,
this defines the critical flux $\Phi_c$, which defines the boundary between
normal and superconducting states.  We then determine that the critical flux
$\phi_c\equiv\Phi_c/(h/e)$ satisfies
\begin{equation}
\frac{2R\sqrt{\pi}}{\sqrt{\xi_0 l_e}}=2\pi \min_{m\in Z}|2\phi_c -{m}|,
\end{equation}
which gives the critical flux for a given radius, or vice versa, via
\begin{equation}
\cos\left( \frac{R}{\sqrt{\xi_0 l_e}/2\sqrt{\pi}}\right)=\cos \left(\frac{2\pi
\Phi_c}{h/2e}\right).
\end{equation}
 When the flux dependence of the
critical temperature is weak, as in the large-radius limit, $t\equiv 1-
T_\text{c}/T_\text{c}^0$ is close to unity and $\Gamma
l_e\xi_0x_m^2(\phi)/R^2\ll1$, so one can expand the logarithm in $(1-t)$
 and the second digamma function around $1/2$ in Eq.~(\ref{eqn:bulk_dirty}) to obtain
\begin{equation}
\label{eqn:TcdirtyApprox1} (1-t) \approx \frac{\pi^2}{2}\frac{\Gamma
 l_e\xi_0}{ R^2}(\phi-\frac{m}{2})^2.
\end{equation}
As we did for Eq.~(\ref{eqn:dGclean}) in the clean limit, we can express this
equation as
\begin{equation}
\label{eqn:deG} \cos\left(2\pi\frac{R}{\xi(t)}\right)\approx \cos\left(2\pi\frac{\Phi}{h/2e}\right),
\end{equation}
where $\xi(t)\approx 0.84\Gamma \sqrt{\xi_0 l_e}/\sqrt{1-t}$ is the
temperature-dependent Ginzburg-Landau coherence length in the dirty limit.
Thus, we recover the de Gennes results~\cite{deGennes}, but via a microscopic
calculation. We note that, strictly speaking, the Ginzburg-Landau approach is
only valid near $t\approx 1$ (see also discussion at the end of
Sec.~\ref{sec:LargeRadiusClean}).

\subsubsection{Finite-radius regime}
Strictly speaking, the de Gennes results~(\ref{eqn:deG}) are only valid in the
large-radius limit, as is clearly seen from our microscopic derivation. In the
regime in which the radius of the ring is comparable to the zero-temperature
coherence length or the Cooper-pair size, we should take into account the
effect of finite radius and use Eq.~(\ref{eqn:finite_dirty}) instead of
Eq.~(\ref{eqn:bulk_dirty}). In addition to the lengthscale defined by the
Cooper-pair size $\xi_0$, the behavior of $T_\text{c}$ also depends on another
length scale, viz., the mean-free path $l_e$.  The correction terms (i.e.~the
terms in the summation) in Eq.~(\ref{eqn:finite_dirty}) decay exponentially
with $R/l_e$, so that the series converges rather rapidly, even for $R\sim
l_e$ or slightly smaller. In Fig.~\ref{fig:dirty} we contrast the predictions
of $T_\text{c}$ in the large-radius limit, Eq.~(\ref{eqn:bulk_dirty}), to
those that include the finiteness corrections, Eq.~(\ref{eqn:finite_dirty}),
for a small radius ($R\sim 0.2\xi_0,l_e$). Figure~\ref{fig:dirty} also shows
that the
 $T_\text{c}(\phi)$ oscillation has a component of period $h/e$, in
addition to the usual Little-Parks component of period $h/2e$. In
Fig.~\ref{fig:Phasedirty} we plot the critical temperature vs.~radius for
various flux values. We see that the deviation from the bulk result is mroe
significant near $\phi=1/2$, whereas the bulk result shows almost no deviation
in the range $\phi=0-0.25$.

The deviation of the critical temperature of small-radius rings from that of
large-radius rings is, however, not universal, as it depends on both the
magnitude and sign of $\cos(2\pi\overline{n})=\cos(2\pi \sqrt{2MR^2\mu})$. We
have seen that when the sign of $\cos(2\pi \sqrt{2MR^2\mu})$ is positive, the
corrections from the finiteness of the radius can cause a reduction in the
critical temperature near the flux value $\Phi=h/2e$, compared to the bulk
case. This corresponds to the case in which, at zero flux, all the electrons
are paired up. However, when the sign of the cosine is negative, the critical
temperature at $\Phi=h/2e$ can actually exceed its value at zero flux; this
corresponds to the case in which, at $\Phi=h/2e$, all electrons are paired up,
thus the system is more stable there than at $\Phi=0$. If the cosine term
should happen to vanish accidentally, the oscillation at period $h/e$ would
disappear.

 When the radius is not small the change in $T_\text{c}(\phi)$, relative to $T_\text{c}(0)$, is small,
 and we
 therefore need only retain terms in Eq.~(\ref{eqn:finite_dirty}) to order $k=1$ and also may replace $t$ on the r.h.s.~by 1, as we did to obtain Eq.~(\ref{eqn:TcdirtyApprox1}).
Thus we obtain
\begin{eqnarray}
(1-t) \approx \frac{\pi^2}{2}\frac{\Gamma
 l_e\xi_0}{ R^2}(\phi-\frac{m}{2})^2 +4
e^{-\pi\frac{R}{l_e}} \cos(2\pi\overline{n})\tanh^{-1}\big(e^{-\frac{2\pi
 }{\Gamma}\frac{R}{\xi_0}}\big) (1-\cos2\pi  \phi).
\end{eqnarray}
Hence, an oscillation of period $h/e$ is clearly seen to emerge, and whether
the critical temperature reduces or increases (e.g.~at $\Phi=h/2e$), compared
to the bulk Little-Parks value, is seen to depend on the sign of the cosine
term. For the case where all electrons are paired up at zero flux, the
critical temperature is lower at $\Phi=h/2e$ than at $\Phi=0$.

\section{Concluding remarks}
\label{sec:conclude}
 We have considered the oscillations in the critical temperature of a superconducting ring
 of finite radius
 in the presence of a threading magnetic flux. We have found that, as the radius of the ring
is (parametrically) reduced, an oscillation in the critical temperature of
period of $h/e$ emerges, in addition to the usual Little-Parks dependence (the
period of which is $h/2e$). Our results provide  corrections, due to the
finiteness of the  ring radius, to the results that de Gennes obtained for a
flux-threaded ring~\cite{deGennes}. We have argued that in the clean limit
there is a superconductor-normal transition, as the ring radius becomes
sufficiently small at nonzero flux, and that the transition can be either
continuous or discontinuous, depending on the radius and/or flux. In the
disordered regime, we have argued that the transition is rendered continuous,
which results in a quantum critical point tuned by flux and radius.

One may wonder how the system behaves as it goes from clean to dirty limit. At
which point does the existence of multiple solutions in the critical
temperature disappear? By analyzing Eq.~(\ref{eqn:arbitraryDisorder}), we
obtain that, ignoring the finiteness corrections, double solutions disappear
when the disorder is such that $l_e/\xi_0\lesssim 1.73$.

One question we should also address is the thickness $d$ of the ring
cross-section. It causes an orbital pair-breaking effect. For the purpose of
estimating this we can use the result from a wire with same thickness in a
field perpendicular to the wire axis (for the calculation, see,
e.g.~Ref.\cite{Tinkham}). The fractional decrease of critical temperature,
when it is small, can be estimated to be (not including the oscillation by
flux)
\begin{equation}
1-t(\phi)\approx \frac{\pi^2\Gamma}{24} \phi^2 \frac{\xi_0}{l_e} \frac{l_e^2
d^2}{R^4},
\end{equation}
which causes a quadratic decrease in the critical temperature on top of the
oscillations discussed in the present paper. If we take the same values shown
in Fig.~\ref{fig:dirty}, i.e., $R\approx l_e\approx 0.2 \xi_0$, for $d\approx
0.2 R$ the decrease is about $3.6\%$ at $\phi=1/2$ and about $14.4\%$ at
$\phi=1$. The largest decrease in the critical temperature shown in
Fig.~\ref{fig:dirty} is about $20\%$, and the $h/e$ component is not swamped
by the pair-breaking effect and can still be observed.

Although we have obtained the emergence of the $h/e$ oscillation and its
amplitude by the microscopic calculations, the physics behind these effects
can be illustrated via heuristic arguments associated with  the corresponding
Cooper problem on a ring, together with a path-integral based  instanton
tunneling approach. We briefly discuss these points in
Appendix~\ref{app:Heuristic}.

The non-universal factor
$\cos(2\pi\overline{n})=\cos\big(2\pi\sqrt{2MR^2\mu}\big)$ determines whether
the finiteness of the ring radius leads to a decrease or an increase in the
critical temperature near flux $\Phi=h/2e$. Although this factor is model
dependent (i.e. dependent on the form of the single-particle spectrum), for
the quadratic spectrum we consider here, $\sqrt{2MR^2\mu}$ roughly counts the
number of electrons divided by four, as there are two spin species and
positive and negative (angular) momenta. If we restrict ourselves to the case
in which all electrons are paired then $\overline{n}=\sqrt{2MR^2\mu}\approx
(N_{\rm pair}-1)/2$, because for $\mu=0$ two electrons can still occupy the
$n=0$ state. When all the levels at $\mu$ and below are filled at zero flux,
$\cos\big(2\pi\sqrt{2MR^2\mu})\approx 1$. When there is one pair fewer or more
(or equivalently, when all pairs are occupied at $\Phi=h/2e$),
$\cos\big(2\pi\sqrt{2MR^2\mu})\approx -1$. This seems to result in an even-odd
effect, not from the number of electrons~\cite{RalphBlackTinkham95} but from
the number of Cooper pairs. However, whether this even-odd effect holds in
general requires further investigation.

\medskip
{\it Notes added in proofs}.  1. We would like to point out that the issue of
flux-dependent supercurrents for s-wave rings was also studied by Zhu and Wang
in Ref.~\cite{28}. In addition to
Refs.~\cite{Dwave,JuricicHerbutTesanovic07,Barash07}, there is a more recent
work on the same issue in d-wave superconductors in Ref.~\cite{29}.

2. Throughout our paper, we have essentially assumed that the switching of
pairing configuration (see Fig. 2) occurs at flux values being an odd integer
multiple of $h/4e$ near superconducting-normal transition temperatures. This
is consistent with our numerical results that by allowing the switching to be
varied, the largest possible critical temperature is obtained when the above
assumption is obeyed.  In a very recent paper by Vakaryuk in Ref.~\cite{30},
the author concludes that the switching can be flux dependent at zero
temperature. If the two results are to be consistent, we are led to the
conclusion that the switching is temperature dependent. This requires further
investigation.

\medskip
 \noindent {\it Acknowledgments\/}. The authors
acknowledge informative discussions with David Pekker and Frank Wilhelm, and
are especially grateful to Gianluigi Catelani for his critical reading of this
paper. This work was supported by DOE Grant No. DEFG02-91ER45439 through the
Federick Seitz Material Research Laboratory at the University of Illinois. TCW
acknowledges the financial support from IQC, NSERC, and ORF, as well as the
hospitality of Perimeter Institute, where part of this work was done. We would
also like to acknowledge V. Vakaryuk and A. J. Leggett for communicating the
results of Ref.~\cite{30}.

\appendix
\section{Averaged Gor'kov equations over radial direction}
\label{app:Gorkov} Consider a ring on a plane with inner and outer radii $a$
and $A$, respectively. Assume that the thickness $(A-a)$ is much smaller than
the mean radius $(A+a)/2$ and the zero-temperature coherence length $\xi(0)$.
We reduce the two dimensional Gor'kov equations into effective one-dimensional
equations by defining the averaged Green function,
\begin{equation}
\widetilde{G}(\theta,\theta')\equiv\frac{2}{A^2-a^2}\int_{a}^A d\rho \rho
\int_{a}^{A}d\rho' \rho' G(\rho,\theta,\rho',\theta';\omega),
\end{equation}
and similarly for $\widetilde{F}^\dagger$. Under this averaging, the
two-dimensional delta function becomes one-dimensional,
\begin{equation}
\frac{2}{A^2-a^2}\int_{a}^A d\rho \rho \int_{a}^{A}d\rho' \rho'
\delta(\vec{r}-\vec{r}')= \delta (\theta-\theta').
\end{equation}
In the Laplacian operator, there is a term
$\frac{1}{\rho}\frac{\partial}{\partial\rho}{\rho}\frac{\partial}{\partial\rho}
G(\rho,\rho')$. When performing the above average, we get for this term
\begin{equation}
\frac{2}{A^2-a^2}\int_{a}^A d\rho \rho \int_{a}^{A}d\rho' \rho'
\frac{1}{\rho}\frac{\partial}{\partial\rho}{\rho}\frac{\partial}{\partial\rho}
G(\rho,\theta,\rho',\theta';\omega)=\frac{2}{A^2-a^2} \int_{a}^{A}d\rho'
\rho'\rho\frac{\partial}{\partial\rho}G\Big\vert_{\rho=a}^{\rho=A},
\end{equation}
which, under the condition of no current flowing radially through the ring boundary, gives zero
identically. Similarly, if we regard the anomalous Green function as playing
the role of the Ginzburg-Landau order parameter in the theory of
superconductivity, no-flow of supercurrent radially gives the average to be
zero for $F^\dagger$. Furthermore, under the limit that the average of the
product of two functions can be approximated by the product of two averaged
functions, we obtain a set of reduced one-dimensional Gor'kov equations, with
the coordinate variables being azimuthal angles $\theta$ and $\theta'$ and the
radii being set to the fixed value of the average radius. After appropriately
renormalizing the Green functions by the inverse of the average radius (thus
making the dimension consistent), we arrive at Eqs.~(\ref{eqn:Gorkov1D}). We
remark that even if the ring has the geometry of a torus, the same averaged 1D
equations will result provided that the cross-section is relatively small, compared with the ring
area and the coherence length.

\section{Critical temperature equations: supplementary details}
\label{app:Detail} In this appendix, we supplement the details leading to the
critical temperature equations in both clean limit and the disordered regime.
\subsection{Clean limit}
\label{app:clean}
 Continuing from Eq.~(\ref{eqn:PoissonSum}), we make a shift in
$x$: $x\rightarrow x - m/2$, under which the integral becomes
\begin{equation}
\label{eqn:doubleG}
 \sum_{k \in Z} \int_{-\inf}^{\inf} dx \frac{e^{-i\pi m
k}e^{i2\pi x
k}}{[i\omega_n-\frac{1}{2MR^2}(x-x_0)^2+\mu][-i\omega_n-\frac{1}{2MR^2}(x+x_0)^2+\mu]},
\end{equation}
where $x_0\equiv \phi-m/2$. The integral over $x$ can be performed using
contour integration. For example, for $k\ge 0$ and $\omega_n>0$, we can close
the contour in the upper plane and evaluate the residues at
$x=x_0+\sqrt{2MR^2(\mu+i\omega_n)}$ and $x=-x_0-\sqrt{2MR^2(\mu-i\omega_n)}$.
We end up with (noting the cancellation of the factor $e^{-i\pi m k}$)
\begin{eqnarray}
&&\frac{\pi e^{i2\pi k \phi+i2\pi k\sqrt{2MR^2(\mu+i\omega_n)}
}}{\sqrt{\frac{2}{MR^2}(\mu+i\omega_n)}\Big[\omega_n+\frac{i x_0^2}{MR^2}-i
x_0\sqrt{\frac{2}{MR^2}(\mu+i\omega_n)}\,\Big]}\cr
 &&\quad + \frac{\pi e^{-i2\pi k
\phi-i2\pi k\sqrt{2MR^2(\mu-i\omega_n)}
}}{\sqrt{\frac{2}{MR^2}(\mu-i\omega_n)}\Big[\omega_n-\frac{i x_0^2}{MR^2}+i
x_0\sqrt{\frac{2}{MR^2}(\mu-i\omega_n)}\,\Big]}.
\end{eqnarray}
Including the contributions from the three other cases, we arrive at the
self-consistency equation:
\begin{eqnarray}
1&=&\frac{|V|T}{2\pi R}{\rm
Re}\sum_{\omega_n>0}^{\omega_D}\left\{\frac{4\pi}{\sqrt{\frac{2}{MR^2}(\mu+i\omega_n)}\Big[\omega_n+\frac{i
x_0^2}{MR^2}-i x_0\sqrt{\frac{2}{MR^2}(\mu+i\omega_n)}\,\Big]}\right.\cr
&&\left.\quad+\sum_{k=1}^\infty\frac{8\pi\, e^{i2\pi
k\sqrt{2MR^2(\mu+i\omega_n) }}\cos2\pi
k\phi}{\sqrt{\frac{2}{MR^2}(\mu+i\omega_n)}\Big[\omega_n+\frac{i
x_0^2}{MR^2}-i x_0\sqrt{\frac{2}{MR^2}(\mu+i\omega_n)}\,\Big]} \right\},
\end{eqnarray}
where we have put the Debye frequency $\omega_D$ as the upper cutoff in the
Matsubara sum. This equation can be re-written as
\begin{eqnarray}
1&=&\sqrt{\frac{2M}{\mu}}|V|T\,{\rm
Re}\sum_{\omega_n>0}^{\omega_D}\left\{\frac{1}{\sqrt{1+i\frac{\omega_n}{\mu}}\Big[\omega_n+\frac{i
x_0^2}{MR^2}-i x_0\sqrt{\frac{2}{MR^2}(\mu+i\omega_n)}\,\Big]}\right.\cr
&&\left.\quad+\sum_{k=1}^\infty\frac{2 e^{i2\pi k\sqrt{2MR^2(\mu+i\omega_n)
}}\cos2\pi k\phi}{\sqrt{1+i\frac{\omega_n}{\mu}}\Big[\omega_n+\frac{i
x_0^2}{MR^2}-i x_0\sqrt{\frac{2}{MR^2}(\mu+i\omega_n)}\,\Big]} \right\}.
\label{eqn:SelfConsistent}
\end{eqnarray}
Furthermore, for typical temperatures we have $\sqrt{\mu+i\omega_n}\approx
\sqrt{\mu}(1+i\omega_n/2\mu)$. The second term causes the contribution of the
$k\ne 0$ terms to be exponentially small for large $R$, i.e.~ there is a factor
\begin{equation}
e^{-2\pi k R\sqrt{2M\mu}\,\omega_n/2\mu}.
\end{equation}

\subsection{Large-radius limit}
\label{app:cleanlarge}
 For sufficiently large $R$, we can ignore the $k\ne 0$
correction terms, in which case we have
\begin{eqnarray}
1&=&\sqrt{\frac{2M}{\mu}}|V|T\,{\rm
Re}\sum_{\omega_n>0}^{\omega_D}\left\{\frac{1}{\sqrt{1+i\frac{\omega_n}{\mu}}\Big[\omega_n+\frac{i
x_0^2}{MR^2}-i x_0\sqrt{\frac{2}{MR^2}(\mu+i\omega_n)}\,\Big]} \right\}.
\end{eqnarray}
The third term in the denominator is usually much larger than the second term,
as the chemical potential $\mu$ is much larger than the level spacing
$1/2MR^2$, i.e., $\sqrt{2MR^2\mu} \gg 1$. For typical values of $\omega_D$ and
$\mu$, we have $\omega_D/\mu\ll 1$. Hence, we have
\begin{eqnarray}
\label{eqn:bulkselfconsistency}
1&\approx&\sqrt{\frac{2M}{\mu}}|V|T\,{\rm
Re}\sum_{\omega_n>0}^{\omega_D}\frac{1}{\omega_n-i x_0\sqrt{\frac{2\mu}{MR^2}}} \\
&=&\sqrt{\frac{2M}{\mu}}\frac{|V|}{2\pi}\,{\rm Re}\sum_{n=0}^{\omega_D/2\pi
T}\frac{1}{(n+\frac{1}{2})- \frac{i x_0}{2\pi T}\sqrt{\frac{2\mu}{MR^2}}
}\label{eqn:secondOfselfconsistencyBulk}.
\end{eqnarray}
The solution for $T$ to this equation gives the critical temperature
$T_\text{c}$. Denoting by $T^0_\text{c}$ the critical temperature in the absence of flux (so that $x_0=0$),
we have the corresponding equation
\begin{eqnarray}
\label{eqn:thirdOfselfconsistencyBulk}
1&\approx&\sqrt{\frac{2M}{\mu}}\frac{|V|}{2\pi}\,\sum_{n=0}^{\omega_D/2\pi
T^0_\text{c}}\frac{1}{n+\frac{1}{2} }.
\end{eqnarray}
Taking the difference of  equations~(\ref{eqn:bulkselfconsistency}) and~(\ref{eqn:thirdOfselfconsistencyBulk}), we have
\begin{eqnarray}
0={\rm Re}\sum_{n=0}^{\omega_D/2\pi T_\text{c}(\phi)}\frac{1}{(n+\frac{1}{2})-
\frac{i x_0}{2\pi T_\text{c}(\phi)}\sqrt{\frac{2\mu}{MR^2}}
}-\sum_{n=0}^{\omega_D/2\pi T^0_\text{c}}\frac{1}{n+\frac{1}{2} }.
\end{eqnarray}
If we extend both upper limits to infinity, we should compensate by the
difference (assuming $\omega_D/T\gg 1$), i.e.,
\begin{equation}
\label{eqn:log}
 \sum_{\big(\omega_D/2\pi T_\text{c}(\phi)\big)+1}^{\big(\omega_D/2\pi
T^0_\text{c}\big)+1}\frac{1}{n+\frac{1}{2}}\approx
\ln\left(\frac{T_\text{c}(\phi)}{T^0_\text{c}}\right).
\end{equation}
Therefore, we arrive at
\begin{eqnarray}
\label{eqn:diff}
 0={\rm Re}\sum_{n=0}^{\infty}\left(\frac{1}{(n+\frac{1}{2})-
\frac{i x_0}{2\pi T_\text{c}(\phi)}\sqrt{\frac{2\mu}{MR^2}}
}-\frac{1}{n+\frac{1}{2}
}\right)+\ln\left(\frac{T_\text{c}(\phi)}{T^0_\text{c}}\right).
\end{eqnarray}
In terms of the digamma function $\psi(x)$ in Eq.~(\ref{eqn:digamma}), we have an implicit formula for $T_\text{c}(\phi)$:
\begin{eqnarray}
\label{eqn:TcBulk2}
\ln\frac{T_\text{c}(\phi)}{T^0_\text{c}}=\psi\left(\frac{1}{2}\right)-{\rm
Re}\,\psi\left(\frac{1}{2}- \frac{i x_0}{2\pi
T_\text{c}(\phi)}\sqrt{\frac{2\mu}{MR^2}} \right).
\end{eqnarray}
Using the quantities $\mu\approx v_F^2/2M$, $\xi_0=v_F/\pi\Delta_0$, and
$\rho\approx R/\xi_0$,  and defining $t(\phi)\equiv T_\text{c}(\phi)/T^0_\text{c}$, we can
re-write Eq.~(\ref{eqn:TcBulk2}) as
\begin{eqnarray}
\ln t(\phi)=\psi\left(\frac{1}{2}\right)-{\rm Re}\,\psi\left(\frac{1}{2}-
\frac{i x_0 \Gamma}{2\rho\, t(\phi)} \right).
\end{eqnarray}

\subsection{Finiteness correction}
\label{app:cleanfinite} What is the correction to $T_\text{c}(\phi)$ due to the finiteness of the radius? To address this question we need
to take into account the $k\ne 0$ corrections to Eq.~(\ref{eqn:SelfConsistent}).
If we take  $\omega_D/\mu\ll 1$ and $\sqrt{\mu+i\omega_n}\approx
\sqrt{\mu}(1+i\omega_n/2\mu)$, the self-consistency
equation~(\ref{eqn:SelfConsistent}) can be approximated as
\begin{eqnarray}
\!\!\!\!\!\!\!\!\!\!\!\!\!\!\!\!\!1&\approx&\sqrt{\frac{2M}{\mu}}\frac{|V|}{2\pi}{\rm
Re}\sum_{n=0}^{\omega_D/2\pi T}\frac{1}{(n\!+\!\frac{1}{2}) - \frac{i x_0
\Gamma T^0_\text{c} }{2\rho\, T} }\Big\{1+ \sum_{k=1}^\infty{2 e^{i2\pi
k\sqrt{2MR^2(\mu+i\omega_n) }}\cos2\pi k\phi}\Big\}\\
\!\!\!\!\!\!\!\!\!\!\!\!\!\!\!\!\!&\approx&\sqrt{\frac{2M}{\mu}}\frac{|V|}{2\pi}{\rm
Re}\sum_{n=0}^{\omega_D/2\pi T}\frac{1}{(n\!+\!\frac{1}{2}) - \frac{i x_0
\Gamma T^0_\text{c} }{2\rho\, T} }\Big\{1+ \sum_{k=1}^\infty{2 e^{i2\pi
k\sqrt{2MR^2\mu} - k\frac{4\pi\rho T}{\Gamma T^0_\text{c}}(n+\frac{1}{2})
}\cos2\pi k\phi}\Big\}.
\end{eqnarray}
Taking the difference between this equation and the version corresponding to $\phi=0$, and
using the trick for converting the cutoff at the Debye frequency in a logarithm (see Eqs.~(\ref{eqn:log})-(\ref{eqn:TcBulk2}),
we arrive at
\begin{eqnarray}
&&\!\!\!\!\!\!\ln t(\phi)=\psi\Big(\frac{1}{2}\Big)-{\rm
Re}\,\psi\Big(\frac{1}{2}\! -\! \frac{i x_0 \Gamma}{2\rho\, t}  \Big) -
4\sum_{k=1}^\infty \left\{ \cos\big(2\pi
 k\sqrt{2MR^2\mu}\big) \,e^{-\frac{2\pi k\rho }{\Gamma}} {\rm HF}
\big[\frac{1}{2},1,\frac{3}{2},e^{-\frac{4\pi k\rho }{\Gamma}}\big]\right. \cr
&&\!\!\!\!\!\!\!\!\!\!\left. -  {\rm Re}\Big\{e^{i2\pi k\sqrt{2MR^2\mu}}
e^{-\frac{2\pi k\rho\, t}{\Gamma}} {\rm HF} \big[\frac{1}{2}\!-\! \frac{i x_0
\Gamma}{2\rho\, t(\phi)},1,\frac{3}{2}\!-\! \frac{i x_0 \Gamma}{2\rho\,
t},e^{-\frac{4\pi k\rho\, t}{\Gamma}}\big]\cos2\pi k\phi\big/\big(1\!-\!
\frac{i x_0 \Gamma}{\rho\, t}\big)\Big\}\right\},
\end{eqnarray}
where the $\phi$ dependence of $t$ on r.h.s. is suppressed. We have extended
the upper limit of the sum over $n$ to infinity for the exponentially decaying
terms (which introduces a negligible small error), and we have used the
formula
\begin{equation}
\label{eqn:sumhypergeometric}
\sum_{n=0}^\infty\frac{e^{-b(n+\frac{1}{2})}}{n+\frac{1}{2}+a}=2 e^{-b/2} \,{\rm
HF}\big[\frac{1}{2}+a,1,\frac{3}{2}+a,e^{-b}\big]\Big/(1+2a),
\end{equation}
where ${\rm HF}[a, b, c, z]$ is the hypergeometric function ${}_2F_1(a, b; c;
z)$.
\subsection{Disordered regime}
\label{app:dirtylimit}

 In this subsection, we  calculate $\widetilde{G}_\omega(n)$ and $K_\omega$
and obtain the r.h.s. of the self-consistency equation~(\ref{eqn:Kernel}). For
simplicity, we  assume that the potential $u(r)$ is short-ranged, so that its
Fourier transform $u(q)$ can be treated as a constant, essentially independent
of momentum transfer $q$.

The one-particle self-energy can be obtained from summing
one-particle-irreducible diagrams (and ignoring the crossed diagrams), and thus we
obtain
\begin{equation}
\Sigma(\omega)=\sum_{n_1}\frac{n_{\rm imp}|u|^2}{i\omega
-\frac{1}{2MR^2}(n_1-\phi)^2+\mu}\,\,\approx -\frac{i}{2\tau(\omega)}{\rm
sgn}(\omega),
\end{equation}
 the real part of the self-energy has been ignored and we shall call
$\tau(\omega)$ the frequency-dependent scattering time. Under the condition that
the Debye frequency is much smaller than the chemical potential, i.e.~$\omega_D/\mu
\ll 1$, and hence for the range of $\omega$'s that are relevant to
superconductivity (i.e.~$|\omega|<\omega_D$), we obtain
\begin{equation}
\frac{1}{2\tau(\omega)}\approx\frac{1}{2\tau_0}\Big[1+\sum_{k>0} 2\cos(2\pi
k\phi) \cos[2\pi k\sqrt{2MR^2\mu}] \,e^{-2\pi k
\sqrt{\frac{MR^2}{2\mu}}|\omega|}\Big],
\end{equation}
where
\begin{equation} \label{eqn:tau0}
 \frac{1}{2\tau_0}\equiv \frac{2\pi n_{\rm
imp}|u|^2}{\sqrt{{2\mu}/{MR^2}}}.
\end{equation}
This form for $\tau(\omega)$ is an approximant because it is calculated with
$G^0$ rather than $\widetilde{G}$. To improve the approximation we then use
the exact disordered Green function $\widetilde{G}$ to calculate the same
self-energy diagrams again (see, e.g. Ref.~\cite{AGD}), thus obtaining a
self-consistency condition for $\tau(\omega)$ (that can be solved
iteratively):
\begin{equation}
\frac{1}{2\tau(\omega)}=\frac{1}{2\tau_0}\Big[1+\sum_{k>0} 2\cos(2\pi k\phi)
\cos[2\pi k\sqrt{2MR^2\mu}] \,e^{-2\pi k
\sqrt{\frac{MR^2}{2\mu}}|\omega|\big(1+\frac{1}{2\tau(\omega)|\omega|}\big)}\Big].
\end{equation}
 As the $k>0$ terms are exponentially
small, we can approximate $\tau(\omega)$ on the right-hand side by $\tau_0$ to
arrive at
\begin{equation}
\label{eqn:tau1}
\frac{1}{2\tau(\omega)}\approx\frac{1}{2\tau_0}\Big[1+\sum_{k>0} 2\cos(2\pi
k\phi) \cos[2\pi k\sqrt{2MR^2\mu}] \,e^{-2\pi k
\sqrt{\frac{MR^2}{2\mu}}|\omega|\big(1+\frac{1}{2\tau_0|\omega|}\big)}\Big].
\end{equation}

Returning to the kernel ${K}$, we re-write Eq.~(\ref{eqn:Kernel}), following
Gor'kov~\cite{Gorkov}, as
\begin{equation}
\label{eqn:K}
 K_\omega(n_1,n_2)=G_\omega(n_1)\, G_{-\omega}(n_2)\big[1+ {\cal
L}_\omega(n_1+n_2)\big],
\end{equation}
where  we have conveniently dropped the \,$\widetilde{}$\, sign in
$\widetilde{G}$, and
\begin{equation}
\label{eqn:L}
 {\cal L}_\omega(n_1+n_2)\equiv n_{\rm imp}\sum_q |u(q)|^2
K_\omega(n_1-q,n_2+q)
\end{equation}
and the argument $n_1+n_2$ in ${\cal L}$ reflects the conservation of total
(azimuthal) momentum of the two incoming (and outgoing) electrons in the
ladder diagrams after disorder averaging. For the purpose of evaluating
self-consistency equation~(\ref{eqn:Self-consistency1}), our goal is to obtain
$K_\omega$ for $n_2=-n_1+m$. By eliminating ${K}$ from Eqs.~(\ref{eqn:K})
and~(\ref{eqn:L}) by we have an equation for ${\cal L}_\omega$:
\begin{equation}
{\cal L}_\omega(m)=n_{\rm imp} \sum_{n_1'}|u|^2
G_\omega(n_1')G_{-\omega}(-n_1'+m)\big(1+{\cal L}_\omega(m)\big),
\end{equation}
which gives
\begin{equation}
{\cal L}_\omega(m)=\frac{A_\omega(m)}{1-A_\omega(m)},
\end{equation}
where
\begin{equation}
{A_\omega}(m)\equiv n_{\rm imp} \sum_{n_1'}|u|^2
G_\omega(n_1')\,G_{-\omega}(-n_1'+m).
\end{equation}
The self-consistency equation~(\ref{eqn:Self-consistency1}) then becomes
\begin{equation}
\label{eqn:Self-consistency2}
 1=\frac{V}{\beta L}\sum_\omega \frac{1}{n_{\rm
imp} |u|^2}\frac{A_\omega(m)}{1-A_\omega(m)}=\frac{V}{\beta
L}\frac{4\pi\tau_0}{\sqrt{2\mu/MR^2}}\sum_\omega
\frac{A_\omega(m)}{1-A_\omega(m)},
\end{equation}
where we have used Eq.~(\ref{eqn:tau0}) in the second equality. As with the
evaluation of Eq.~(\ref{eqn:doubleG}) in the clean limit (see
Appendix~\ref{app:Detail}), we obtain $A_\omega$ as
\begin{eqnarray}
\!\!\!\!\!\!\!\!\!\!\!\!\!\!A_\omega&=&A^0_\omega+A^{k> 0}_\omega \nonumber \\
\!\!\!\!\!\!\!\!\!\!\!\!\!\!&=&\frac{1}{2\tau}{\rm
Re}\,\Big[\frac{1}{|\omega|\eta-i X_0}\Big]+\sum_{k>0}\frac{e^{-2\pi k
\sqrt{\frac{MR^2}{2\mu}}|\omega|\eta}}{\tau}\cos(2\pi k\sqrt{2MR^2\mu}){\rm
Re}\,\Big[\frac{e^{i2\pi k\phi}}{|\omega|\eta-i X_0}\Big],
\end{eqnarray}
where $X_0\equiv (\phi-m/2) \sqrt{\frac{2\mu}{MR^2}}$,
$\eta\equiv(1+2\tau|\omega|)/(2\tau|\omega|)$, and $\tau$ is a shorthand for
$\tau(\omega)$.

To illustrate the corrections arising from the finiteness of the radius, we
re-write $A_\omega$ as
\begin{equation}
\label{eqn:Aab} A_\omega=\frac{(1+2\tau|\omega|)(1+a_\omega)-2\tau b_\omega
X_0}{(1+2\tau|\omega|)^2+(2\tau X_0)^2}\,\,,
\end{equation}
where $a_\omega(\ll 1)$ and $b_\omega (\ll 1)$ are
\begin{subequations}
\begin{eqnarray}
a_\omega&\equiv&\sum_{k>0} 2 {e^{-2\pi k \sqrt{\frac{MR^2}{2\mu}}|\omega|\eta}}\cos(2\pi k\sqrt{2MR^2\mu})\cos(2\pi k \phi),\label{eqn:a}\\
b_\omega&\equiv&\sum_{k>0} 2 {e^{-2\pi k
\sqrt{\frac{MR^2}{2\mu}}|\omega|\eta}}\cos(2\pi k\sqrt{2MR^2\mu})\sin(2\pi k
\phi).
\end{eqnarray}
\end{subequations}
As the correction to $1/\tau(\omega)$ is exponentially small [see
Eq.~(\ref{eqn:tau1})], in the exponentials $\eta$ can be approximated by
\begin{equation}
\label{eqn:EtaApprox} \eta_0\equiv 1+ \frac{1}{2\tau_0|\omega|}.
\end{equation}
On the other hand, for $1/\tau(\omega)$ not in the exponentials we shall
approximate it by retaining the leading correction [see Eqs.~(\ref{eqn:tau1})
and~(\ref{eqn:a})] via
\begin{equation}
\label{eqn:tau}
 \frac{1}{2\tau(\omega)}\approx\frac{1}{2\tau_0}(1+a_\omega).
\end{equation}
The quotient $A_\omega/(1-A_\omega)$ then becomes
\begin{equation}
\label{eqn:A1A}
 \frac{A_\omega}{1-A_\omega}=\frac{1+\big[a_\omega-\frac{2\tau
b_\omega X_0}{1+2\tau|\omega|}\big]}{
2\tau|\omega|+\frac{1}{1+2\tau|\omega|}(2\tau X_0)^2-\big[a_\omega-\frac{2\tau
b_\omega X_0}{1+2\tau|\omega|}\big] }\approx \frac{1+\big[a_\omega-\frac{2\tau
b_\omega X_0}{1+2\tau|\omega|}\big]}{
2\tau|\omega|+\frac{1}{1+2\tau|\omega|}(2\tau X_0)^2} .
\end{equation}
 We remark
that from this equation one can obtain the equation for the critical
temperature for arbitrary mean-free path $l_e$.
\subsubsection{Large-radius limit and arbitrary disorder}
To illustrate the remark made above, we consider large-radius limit in which
we ignore corrections due to the finite radius. In this limit
Eq.~(\ref{eqn:A1A}) becomes
\begin{equation}
 \frac{A_\omega}{1-A_\omega}= \frac{1}{
2\tau_0|\omega|+\frac{1}{1+2\tau_0|\omega|}(2\tau_0 X_0)^2} .
\end{equation}
Substituting this into Eq.~(\ref{eqn:Self-consistency2}) we have
\begin{eqnarray}
\label{eqn:bulkselfconsistencyArbitraryl} 1
&=&\sqrt{\frac{2M}{\mu}}\frac{|V|}{2\pi}\,{\rm Re}\sum_{n=0}^{\omega_D/2\pi
T_\text{c}}\frac{\big(n+\frac{1}{2}\big)+\frac{1}{4\pi\tau_0
T_\text{c}}}{\big(n+\frac{1}{2}\big)\big[\big(n+\frac{1}{2}\big)+\frac{1}{4\pi\tau_0
T_\text{c}}\big]+\Big(\frac{X_0}{2\pi T_\text{c}}\Big)^2
}\label{eqn:secondOfselfconsistencyBulk2}.
\end{eqnarray}
In the clean limit $\tau_0 T_\text{c}\gg 1$, we see that this reduces to
Eq.~(\ref{eqn:secondOfselfconsistencyBulk}), which we obtained in the absence
of disorder. But here $\tau_0$ is arbitrary. Subtracting
Eq.~(\ref{eqn:bulkselfconsistencyArbitraryl}) from the corresponding equation
at $\phi=0$ (i.e.~$X_0=0$) and using the trick to get logarithm of the ratio
of the critical temperatures as we did in
Eqs.~(\ref{eqn:log})-(\ref{eqn:TcBulk2}), we have
\begin{eqnarray}
\ln\left(\frac{T_\text{c}(\phi)}{T_\text{c}^0}\right)
&=&\sum_{n=0}^{\infty}\left(\frac{\big(n+\frac{1}{2}\big)+\frac{1}{4\pi\tau_0
T_\text{c}}}{\big(n+\frac{1}{2}\big)\big[\big(n+\frac{1}{2}\big)+\frac{1}{4\pi\tau_0
T_\text{c}}\big]+\Big(\frac{X_0}{2\pi
T_\text{c}}\Big)^2}-\frac{1}{n+\frac{1}{2}}\right).
\end{eqnarray}
By making the partial fractions of the first term in the summation and by
using the definition of the digamma function~(\ref{eqn:digamma}), we arrive at
\begin{eqnarray}
\label{eqn:arbitraryDisorder}
\ln\left(\frac{T_\text{c}(\phi)}{T_\text{c}^0}\right)
&=&\psi\left(\frac{1}{2}\right)-\frac{1}{\sqrt{\alpha^2-\chi^2}}
\Big[\frac{-\alpha+\sqrt{\alpha^2-\chi^2}}{2}\psi\left(\frac{1+\alpha+\sqrt{\alpha^2-\chi^2}}{2}\right)
\nonumber \\
&&\qquad+\frac{\alpha+\sqrt{\alpha^2-\chi^2}}{2}\psi\left(\frac{1+\alpha-\sqrt{\alpha^2-\chi^2}}{2}\right)
\Big],
\end{eqnarray}
where, for convenience, we have defined $\alpha\equiv
1/4\pi\tau_0T_\text{c}(\phi)$ and $\chi\equiv X_0/\pi T_\text{c}(\phi)$. This
is the equation for the critical temperature at arbitrary disorder for large
radii of rings.
\subsubsection{Finite-radius corrections and strong disordered regime}
Now we return to the corrections due to the finiteness of the radius. In the
strong disordered limit (i.e.~$\tau_0\,\omega\ll 1$), Eq.~(\ref{eqn:A1A})
becomes
\begin{equation}
\frac{A_\omega}{1-A_\omega}\approx\frac{1}{ 2\tau\big[|\omega|+2\tau
X_0^2\big]  }+ \frac{a_\omega-{2\tau b_\omega X_0}}{ 2\tau\big[|\omega|+2\tau
X_0^2\big]  } .
\end{equation}
This is further approximated, using Eq.~(\ref{eqn:tau}) as
\begin{equation}
\frac{A_\omega}{1-A_\omega}\approx\frac{1}{ 2\tau_0\big[|\omega|+2\tau_0
X_0^2\big]  }+ \frac{2a_\omega-{2\tau_0 b_\omega X_0}}{
2\tau_0\big[|\omega|+2\tau_0 X_0^2\big]  },
\end{equation}
where we have used Eq.~(\ref{eqn:tau}). The first term on the r.h.s. represents the bulk
term and is the Little-Parks term in the dirty limit. The second term takes
into account  the finite radius, and contains flux dependence  in period
$h/e$. The equation for the critical temperature will then contain the digamma
function and hypergeometric function with real arguments, in contrast with the
clean limit. This makes the transition to normal state continuous, as the
solution for the critical temperature is unique and the assumption of vanishing
order parameter used in the linearized self-consistency condition is valid.
Therefore, there can be quantum phase transitions tuned by flux and/or radius.

As $\tau_0|X_0|\ll 1$ in the strong-disorder limit, the ratio
$A_\omega/(1-A_\omega)$ becomes (by ignoring the $b$ term)
\begin{equation}
\label{eqn:quotientA}
\frac{A_\omega}{1-A_\omega}\approx\frac{1}{ 2\tau_0\big[|\omega|+2\tau_0
X_0^2\big]  }+ \frac{2a_\omega}{ 2\tau_0\big[|\omega|+2\tau_0 X_0^2\big]  }.
\end{equation}
From this we can obtain an equation for the critical temperature $T_c$ in the
presence of flux $\Phi$,  as we have done in the clean limit (see
Appendix~\ref{app:cleanfinite}). By inserting  Eq.~(\ref{eqn:quotientA}) into the
self-consistency equation~(\ref{eqn:Self-consistency2}) we obtain
\begin{eqnarray}
\!\!\!\!\!\!1=\sqrt{\frac{2M}{\mu}}\frac{V}{2\pi}\sum_{n=0}^{\frac{\omega_D}{2\pi
T_c}} \frac{1+\sum_{k>0}2 e^{-2\pi k
\sqrt{\frac{MR^2}{2\mu}}\big[\frac{1}{2\tau_0} +2\pi
T_c(n+\frac{1}{2})\big]}\cos(2\pi k \sqrt{2MR^2\mu})\cos2\pi
k\phi}{n+\frac{1}{2}+\frac{\tau_0 X_0^2}{\pi T_c}}.
\end{eqnarray}
Subtracting from this the corresponding $\phi=0$ equation, i.e.,
\begin{eqnarray}
1=\sqrt{\frac{2M}{\mu}}\frac{V}{2\pi}\sum_{n=0}^{\frac{\omega_D}{2\pi T_c^0}}
\frac{1+\sum_{k>0}2 e^{-2\pi k \sqrt{\frac{MR^2}{2\mu}}\big[\frac{1}{2\tau_0}
+2\pi T_c^0(n+\frac{1}{2})\big]}\cos(2\pi k \sqrt{2MR^2\mu})}{n+\frac{1}{2}},
\end{eqnarray}
we obtain the following implicit equation for the critical temperature:
\begin{eqnarray}
0&=&\sum_{n=0}^{\infty}\Big(\frac{1}{n+\frac{1}{2}}-\frac{1}{n+\frac{1}{2}+\frac{\tau_0
X_0^2}{\pi T_c}}\Big)-\sum_{n=\frac{\omega_D}{2\pi
T_c^0}+1}^{\frac{\omega_D}{2\pi T_c}}\frac{1}{n} +\sum_{k>0}2e^{-2\pi k
\sqrt{\frac{MR^2}{2\mu}}\frac{1}{2\tau_0}} \cos(2\pi
k \sqrt{2MR^2\mu})\times\nonumber\\
&&\qquad\sum_{n=0}^{\infty}\left(\frac{e^{-2\pi k \sqrt{\frac{MR^2}{2\mu}}2\pi
T_c^0(n+\frac{1}{2})}}{n+\frac{1}{2}}-\frac{e^{-2\pi k
\sqrt{\frac{MR^2}{2\mu}}2\pi T_c(n+\frac{1}{2})}\cos2\pi
k\phi}{n+\frac{1}{2}+\frac{\tau_0 X_0^2}{\pi T_c}}\right),
\end{eqnarray}
where we have used the fact that when $n\approx \omega_D/2\pi T^0_c$ or
higher, $1/(n+\frac{1}{2}+ x)\approx 1/n$, the sum of the series being
approximated is a logarithm, and we have extended the upper limit of $n$ for the
correction terms to infinity. By using  formula (\ref{eqn:sumhypergeometric})
for the hypergeometric function, we finally arrive at
\begin{eqnarray}
&&\ln\left(\frac{T_c}{T_c^0}\right)=\psi\left(\frac{1}{2}\right)-\psi\left(\frac{1}{2}+\frac{\tau_0
X_0^2}{\pi T_c}\right) +\sum_{k>0}4 e^{-2\pi k \sqrt{\frac{MR^2}{2\mu}}
\frac{1}{2\tau_0}} \cos(2\pi k \sqrt{2MR^2\mu})
\nonumber \\
&& \times\left[ e^{-2\pi k \sqrt{\frac{MR^2}{2\mu}}\pi T_c}\cos2\pi k
\phi\,{{\rm HF}\Big(\frac{1}{2}+\frac{\tau_0 X_0^2}{\pi
T_c},1,\frac{3}{2}+\frac{\tau_0 X_0^2}{\pi T_c},e^{-2\pi k
\sqrt{\frac{MR^2}{2\mu}}2\pi T_c}\Big)}\Big/{\big(1+\frac{2\tau_0
X_0^2}{\pi T_c}\big)}\right.\nonumber\\
&&\left. \qquad\quad-e^{-2\pi k \sqrt{\frac{MR^2}{2\mu}}\pi T_c^0} {\rm
HF}\Big(\frac{1}{2},1,\frac{3}{2},e^{-2\pi k \sqrt{\frac{MR^2}{2\mu}}2\pi
T_c^0}\Big)\right],
\end{eqnarray}
where we recall that $X_0\equiv (\phi-m/2)\sqrt{\frac{2\mu}{MR^2}}$. If we use
$\mu=Mv_F^2/2$, $l_e=v_F \tau_0$, and $\xi_0=v_F/\pi \Delta_0$ (with
$\Delta_0=\Gamma\, T_c^0$), we have
 \begin{eqnarray}
 &&\ln t=\psi\Big(\frac{1}{2}\Big)-\psi\Big(\frac{1}{2}+\frac{\Gamma
 l_e\xi_0}{t R^2}(\phi-\frac{m}{2})^2\Big)
 +\sum_{k>0}4 e^{-\pi k \frac{R}{l_e}}\cos(2\pi k \sqrt{2MR^2\mu})
 \nonumber \\
 && \times\left[ e^{-\frac{2\pi k}{\Gamma}\frac{R}{\xi_0} t }\cos2\pi k \phi\,{ {\rm
 HF}\Big(\frac{1}{2}+\frac{\Gamma l_e\xi_0}{t R^2}(\phi-\frac{m}{2})^2,1,\frac{3}{2}+
 \frac{\Gamma l_e\xi_0}{t R^2}(\phi-\frac{m}{2})^2,e^{-\frac{4\pi
 k}{\Gamma}\frac{R}{\xi_0}t}
\Big)}\times\right.\nonumber\\
 &&\left. \qquad\quad{\Big(1+\frac{2\Gamma l_e\xi_0}{t R^2}(\phi-\frac{m}{2})^2\Big)^{-1}}
 -e^{-\frac{2\pi k}{\Gamma}\frac{R}{\xi_0}} {\rm
 HF}\Big(\frac{1}{2},1,\frac{3}{2},e^{-\frac{4\pi
 k}{\Gamma}\frac{R}{\xi_0}}\Big)\right].
\end{eqnarray}
We can re-write this equation using the relations $\mu=M v_F^2/2$,
$\Delta_0=v_F/\pi\xi_0$ and $l_e=v_F\tau_0$, and thus arrive at
 \begin{eqnarray}
 \label{app:eqn:finite_dirty}
 &&\ln t=\psi\Big(\frac{1}{2}\Big)-\psi\Big(\frac{1}{2}+\frac{\Gamma
 l_e\xi_0}{t R^2}(\phi-\frac{m}{2})^2\Big)
 +\sum_{k=1}^\infty4 e^{-\pi k  \frac{R}{l_e}}\cos(2\pi k \sqrt{2MR^2\mu})
 \nonumber \\
 && \left[ e^{-\frac{2\pi k }{\Gamma}\frac{R}{\xi_0} t }\cos2\pi k \phi\frac{ {\rm
 HF}\big(\frac{1}{2}+\frac{\Gamma l_e\xi_0}{t R^2}(\phi-\frac{m}{2})^2,1,\frac{3}{2}+
 \frac{\Gamma l_e\xi_0}{t R^2}(\phi-\frac{m}{2})^2,e^{-\frac{4\pi k
 }{\Gamma}\frac{R}{\xi_0}t\big)
}}{1+\frac{2\Gamma l_e\xi_0}{t R^2}(\phi-\frac{m}{2})^2}\right.\nonumber\\
 &&\left.\quad \qquad-e^{-\frac{2\pi k }{\Gamma}\frac{R}{\xi_0}} {\rm
 HF}\Big(\frac{1}{2},1,\frac{3}{2},e^{-\frac{4\pi k
 }{\Gamma}\frac{R}{\xi_0}}\Big)\right].
\end{eqnarray}
\subsubsection{Finite-radius and arbitrary disorder}
It is also possible to include corrections from the finiteness of the radius,
and derive the equation for the critical temperature, as we did in the
previous section. The results will contain the r.h.s. of
Eq.~(\ref{eqn:arbitraryDisorder}) as the zeroth order, as well as corrections
due to the finiteness of the radius in terms of hypergeometric functions.
However, the formulas are too cumbersome and we do not list them here.
\section{Heuristic argument to the emergence of $h/e$ period oscillations}
\label{app:Heuristic}
\subsection{Cooper problem on a ring}
Let us consider Cooper's problem on a ring with a flux $\Phi$ threading
through it. The orbital wavefunction can be written in the same form as in
Eq.~(\ref{eqn:GnF}). Assuming that $\psi(x,x)\sim e^{i2\pi m x/L}$, and that
the electron-electron interaction is factorizable, i.e.~$V_{n,n'}=\Lambda U_n
U^*_{n'}$, we can write down the time-independent Schr\"odinger equation in terms of the
wavefunction amplitude $a_n$ as
\begin{equation}
2\epsilon_n(\phi) a_n +\sum_{|n'|>n_F} V_{n,n'} a_{n'}= E a_n,
\end{equation}
where
\begin{equation}
2\epsilon_n(\phi)=\frac{1}{2MR^2}\big[(n+\phi)^2+(m+n-\phi)^2\big],
\end{equation}
and hence we arrive at
\begin{equation}
\frac{1}{\Lambda}=\sum_{n}\frac{1}{E-2\epsilon_n(\phi)}.
\end{equation}
The integer $m$ has to be chosen to minimize the ground state energy. Thus, we
see that the solution is periodic in $\phi$ with period 1 (or in $\Phi$ with
period $h/e$, namely the single particle flux quantum) and that the
Little-Parks period $h/2e$ is exact only in the large-$R$ limit, in which case
the summation over $n$ can be replaced by an integral and hence the period of
$\phi$ is $1/2$. This illustrates the role of the single particle oscillation.
However, the amplitudes of the single-particle and Little-Parks oscillations
have to be evaluated within a microscopic theory, as has been done in the main
text.

\subsection{Flux oscillation: an instanton approach}
 The instanton picture provides us another, heuristic, view of the emergence
 of $h/e$-period oscillations. The argument presented here is intended just to give some intuition. We
  refer to Rajaraman~\cite{Rajaraman} for a pedagogical review of instanton techniques.  A single instanton
 tunneling from $0$ through a potential to $2\pi$ (which is identified as $0$)
 on a circle gives an amplitude
 \begin{equation}
 \lim_{\tau\rightarrow\infty}\langle2\pi|e^{-H\tau}|0\rangle_{(1,0)}=e^{-S_0}
 JK\tau\omega^{1/2}e^{-\omega\tau/2},
 \end{equation}
 where $S_0$ is the classical Euclidean action (in the absence of any flux threading through the circle), $J$ is a Jacobian factor,
 $K$ is a constant independent of $\tau$ as
 $\tau\rightarrow\infty$, and $\omega$ is the harmonic oscillator frequency
 near the bottom of the trapping potential. Now, if there is a flux $\Phi$
 threading the ring, the amplitude will also acquire an additional phase factor
 $e^{i2\pi \phi}$, where $\phi\equiv \Phi/\Phi_0$ and $\Phi_0=h/e$, if the particle
 carries charge $e$. For a charge $2e$ particle, the phase factor would be $e^{i4\pi \phi}$.

We consider a ring size that is large enough that the electrons in a Cooper pair
generally traverse the ring together, and rarely  split so as to wind separately around the ring. The ground state energy
 will give a qualitative estimate of the critical temperature of the associated
 superconductor. The total contribution of the
 amplitude for the instanton associated with the Cooper pair is then
\begin{equation}
 \lim_{\tau\rightarrow\infty}\langle2\pi|e^{-H\tau}|0\rangle=\omega^{1/2}e^{-\omega\tau/2}
 \sum_{n_1,n_2}
 \frac{1}{n_1!n_2!}\big(JK\tau e^{-S_0}\big)^{n_1+n_2}e^{i4\pi(n_1-n_2)\phi},
 \end{equation}
 which gives
\begin{equation}
 \lim_{\tau\rightarrow\infty}\langle2\pi|e^{-H\tau}|0\rangle=\omega^{1/2}e^{-\omega\tau/2}
\exp\Big(
 2JK\tau e^{-S_0}\cos{4\pi\phi}\Big)\sim
 \lim_{\tau\rightarrow\infty}\langle2\pi|E_0\rangle\langle
 E_0|0\rangle e^{-E_0\tau}.
 \end{equation}
 This gives a ground state energy of $E_0=\frac{\omega}{2}-2JK
 e^{-S_0}\cos{4\pi\phi}$, which reveals the period $h/2e$, the
 Little-Parks period in critical temperature. The fact that this gives  a
 dependence on flux being sinusoidal rather than quadratic results from the lack of
 accounting of the other many-body electrons and the existence of a condensate.

Now, occassionally (in the sense of contributing Feynman paths) the electrons in a Cooper pair separate and circumnavigate the ring (relative to one another) before re-associating.
 The contribution of such processes to the ground-state energy can be
 estimated via the instantons and anti-instantons of such events associated with them. A single instanton involving  one
 electron going from $0$ to $2\pi$ has the amplitude
\begin{equation}
 \lim_{\tau\rightarrow\infty}\langle2\pi|e^{-H\tau}|0\rangle_{(1,0)}=-e^{-S_{0;e}}
 J_e K_e\tau\omega_e^{1/2}e^{-\omega_e\tau/2} e^{i2\pi\phi},
 \end{equation}
where the subscript $e$ indicates a single electron rather than a Cooper pair,
and the additional minus sign comes from the exchange of the two electrons.
Summing all the instanton and anti-instanton processes, we arrive at the
contribution to the ground-state energy from the two electrons
\begin{equation}
E_{0;e}=\omega_e+4J_eK_e
 e^{-S_{0;e}}\cos{2\pi\phi},
\end{equation}
which results in the emergence of an $h/e$ contribution to the period of oscillation to
critical temperature. We remark that it is owing to the separation of the lengthscales
and hence time scales that we can separate the  Cooper-pair and
single-electron contributions. The amplitudes of the two oscillations are
related to the respective actions, and for large radius, the amplitude
corresponding to single-particle oscillation is expected to be small, due to
the binding resulting from the attractive interparticle interaction.


\begin{thebibliography}{99}
\bibitem{footnotehe}
In C.G.S unit, the single-particle flux quantum is $hc/e$. We shall set $c=1$
and $\hbar=1$. Even though in this unit the flux quantum becomes $2\pi/e$, we
shall refer to it by $h/e$ instead.
\bibitem{Average} Although
we take the ring to be of infinitesimally thin width, for small but finite
width compared to the radius and the coherence length, this set of equations
can be obtained by averaging over the radial direction and ignoring the
fluctuations; see Appendix A. In our discussions we require that the mean
average level spacing $\hbar v_F/d$ in the transverse direction is much
greater than the superconducting gap $\Delta_0$, so that only the lowest level
in the transverse direction is occupied. This leads to the condition that
$d\ll \xi_0$, where $\xi_0$ is the Cooper-pair size $\xi_0=\hbar v_F/\pi
\Delta_0$.
\bibitem{footnoteSarma} Note that here we assume the equilibrium normal state
does not possess a persistent current. Ignoring the finiteness correction, the
transition to normal follows the curve ABD in Fig.~\ref{fig:sfPhase}. If the
normal state does possess a persistent current, the transition from
superconducting to normal will follow the curve ABE instead. The inclusion of
the finiteness correction only modifies the detail locations of these curves.
\bibitem{footnote} This can be seen from the fact that the function
$f(t)=\ln
t-\psi\left(\frac{1}{2}\right)+\left(\frac{1}{2}+\frac{a}{t}\right)$,
is greater than zero for $t\ge 0$ when $a> \Gamma/4\pi$, whereas it crosses
from negative values to positive at a certain value of $t>0$ when
$a<\Gamma/4\pi$.
\end{thebibliography}
\end{document}